\documentstyle[psfig,epsf]{mn}

\def\lessapprox{\,\raise 0.6ex\hbox{$<$}\kern -0.75em\lower 0.47ex
    \hbox{$\sim$}\,}
\def\largapprox{\,\raise 0.6ex\hbox{$>$}\kern -0.75em\lower 0.47ex
    \hbox{$\sim$}\,}

\begin{document}

\title[Void Hierarchy]
{A hierarchy of voids: Much ado about nothing}
\author[R. K. Sheth \& R. van de Weygaert]
{Ravi K. Sheth$^{1}$ \& Rien van de Weygaert$^2$\\
$^1$Dept. of Physics and Astronomy, University of Pittsburgh, 
    3941 O'Hara St., PA 15260, U.S.A.\\
$^2$Kapteyn Institute, University of Groningen, P.O. Box 800, 
    9700 AV Groningen, The Netherlands \\
\smallskip
Email: rks12@pitt.edu, weygaert@astro.rug.nl
}
\date{Submitted to MNRAS 11 November 2003}

\maketitle

\begin{abstract}
We present a model for the distribution of void sizes and its 
evolution in the context of hierarchical scenarios of gravitational 
structure formation. We find that at any cosmic epoch the voids 
have a size distribution which is well-peaked about a characteristic 
void size which evolves self-similarly in time.  This is in distinct 
contrast to the distribution of virialized halo masses which does 
not have a small-scale cut-off. 

In our model, the fate of voids is ruled by two processes.  
The first process affects those voids which are embedded in larger 
underdense regions:  the evolution is effectively one in which a 
larger void is made up by the mergers of smaller voids, and is 
analogous to how massive clusters form from the mergers of less 
massive progenitors.  
The second process is unique to voids, and occurs to voids which 
happen to be embedded within a larger scale overdensity:  these voids 
get squeezed out of existence as the overdensity collapses around 
them.  It is this second process which produces the cut-off at 
small scales.  

In the excursion set formulation of cluster abundance and evolution,  
solution of the {\it cloud-in-cloud} problem, i.e., counting as 
clusters only those objects which are not embedded in larger 
clusters, requires study of random walks crossing {\it one-barrier}.  
We show that a similar formulation of void evolution requires study of 
a {\it two-barrier} problem: one barrier is required to account for 
{\it voids-in-voids}, and the other for {\it voids-in-clouds}.  
Thus, in our model, the void size distribution is a function of two 
parameters, one of which reflects the dynamics of void formation, 
and the other the formation of collapsed objects.  
\end{abstract}


\section{Introduction}\label{intro}
An overwhelming body of observational and theoretical evidence favours 
the view that structure in the Universe has risen out of a nearly 
homogeneous and featureless primordial cosmos through the process of 
gravitational instability. Almost all viable existing theories for 
structure formation within the context of this framework are 
hierarchical: the matter distribution evolves through a sequence 
of ever larger structures. 

Hierarchical scenarios of structure formation have been successful
in explaining the formation histories of gravitationally bound 
virialized haloes. They provide a basic framework within which more 
intricate aspects of the formation of a wide range of cosmic objects, 
ranging from galaxies to rich clusters, may be investigated.  
In particular, a fully analytical description of the collapse and 
virialization of overdense dark matter halos has been developed.  
The approach, originally proposed by Press \& Schechter~\cite{ps74}, 
and later modified by Epstein~\cite{re83} and Bond et al.~\cite{bcek91}, 
has led to simple and accurate models for the abundance of massive 
haloes which results from hierarchical gravitational clustering.  
This framework has come to be called the {\it excursion set approach}.

The excursion set approach provides a useful framework for thinking 
about the formation histories of gravitationally bound virialized 
haloes in scenarios of hierarchical structure formation.  
It provides analytic approximations for the distribution of halo 
masses, merger rates, and formation times which are quite accurate 
(Lacey \& Cole~1993), and can be extended to provide estimates 
of the distribution of the mass in randomly placed cells 
(Sheth~1998). A key ingredient in the original approach, 
inherited from the pioneering work of Press \& Schechter~\cite{ps74}, 
is the assumption that virialized objects form from a smooth 
spherical collapse. In reality the collapse can be quite different 
from spherical; recent work has shown that ellipsoidal collapse can 
be incorporated into the approach, with reasonable improvements in 
accuracy (e.g. Sheth, Mo \& Tormen~2001).  

Models based on spherical evolution are difficult to 
reconcile with the spatial patterns which characterize the cosmic 
matter distribution.  The observed world of galaxy redshift surveys, 
and the artificial world of numerical simulations of cosmic structure 
formation, are both characterized by filamentary and sheetlike 
structures. Such weblike patterns represent distinctly non-virialized 
structures for which gravitational contraction of initially aspherical 
density peaks has only been accomplished along one or two dimensions. 
At first sight, such weblike configurations would seem to be beyond 
the realm of the idealized excursion-set description. 

\begin{figure*}
 \centering
 \vskip -2.0cm
 \epsfxsize=16.2cm
 \mbox{\hskip 0.8truecm\epsfbox{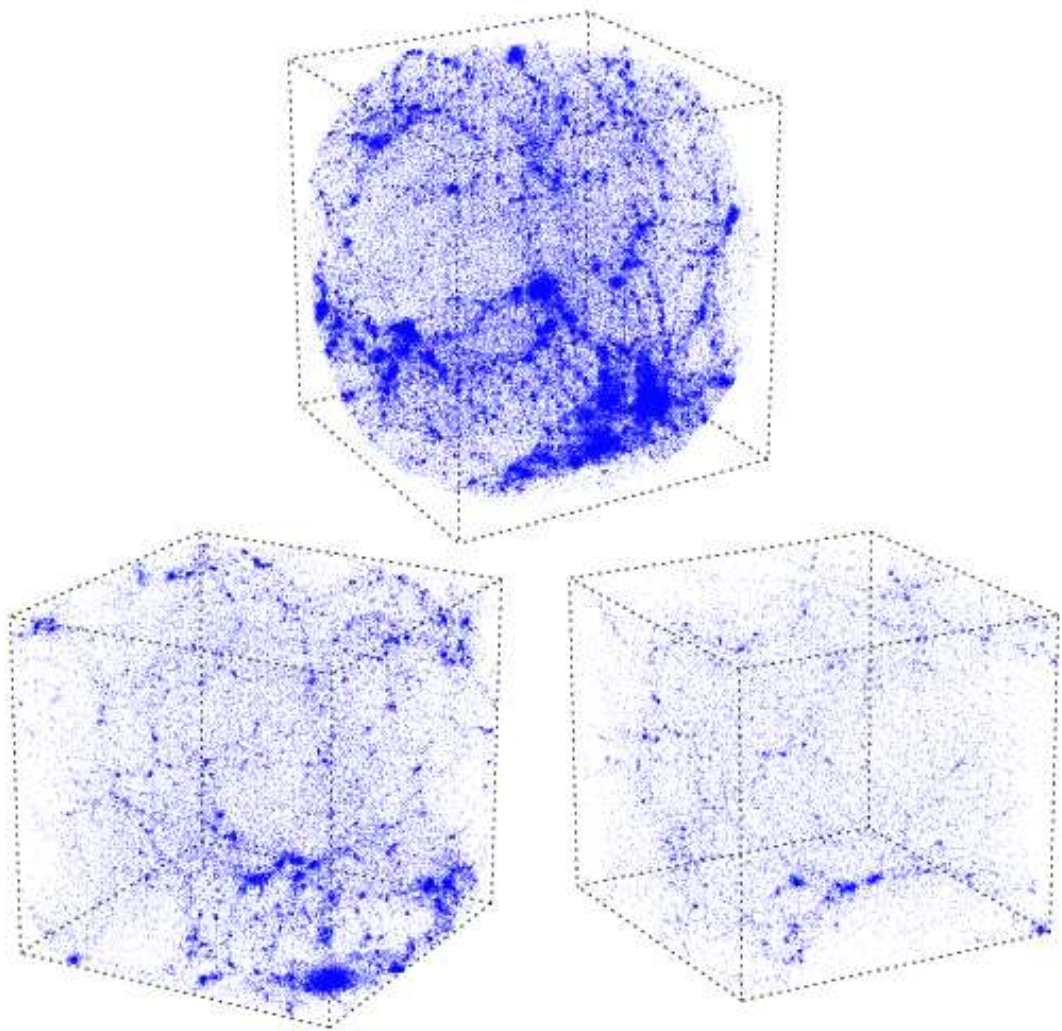}}
 \vskip -1.0cm
 \caption{Spatial structure in a void-like region selected from an 
 N-body simulation of structure formation in the SCDM scenario. 
 Three consecutive zoom-ins centered on the core of the void are 
 shown: a $45\hbox{Mpc}$ diameter particle sphere (top), 
 $36\hbox{Mpc}$ (bottom left) and $30\hbox{Mpc}$ (bottom right).  
 The existence of substructure within the void region is readily 
 apparent, although it becomes more faint and tenuous towards the 
 increasingly depleted interior of the void.}
 \label{voidcdmcube}
\end{figure*}

Nevertheless, in this study we show that the formation and evolution 
of foamlike patterns can indeed be described by the excursion set 
analysis. This is accomplished by focusing on the evolution of 
{\em underdense regions}, the {\em voids}, rather than overdensities 
in the matter distribution. 
Whereas much of the mass in the universe is bound-up in virialized 
structures, most of the volume is occupied by large underdense voids: 
voids are {\it the} dominant component of the Megaparsec-scale galaxy 
and matter distributions. 
In a void-based description of structure formation, 
matter is squeezed in between expanding voids, and sheets and 
filaments form at the intersections of the void walls 
(Icke~1984; Van de Weygaert~1991, 2002). 
Such a view is supported by Reg{\H o}s \& Geller~(1991), 
Dubinski et al.~(1993) and Van de Weygaert \& Van Kampen~(1993), 
who give clear and lucid descriptions of how voids evolve in 
numerical simulations of gravitational clustering. 
We will stick to this basic framework in the present study. 

We will argue that low density regions are the objects-of-choice 
for working out a succesful analytical description of cosmic 
spatial structure, if it is to be based upon the idealization 
of spherical symmetry.  This is because, in many respects, voids are 
ideally suited for an excursion set analysis based on a spherical 
evolution model. 
This is despite the fact that voids form from negative density 
perturbations in the initial fluctuation field, and neither maxima 
nor minima in the primordial Gaussian field, are spherical 
(see Bardeen et al. 1986).  
However, in marked contrast to the evolution of density peaks, 
primordial asphericity of negative density perturbations is 
quickly lost as they expand:  the generic evolution is towards 
an approximately spherical tophat geometry (Icke~1984).  
Moreover, the velocity structure of uniform density voids is simple 
to understand; an observer in the interior will observe a Hubble-type 
velocity field. All of this is discussed in some detail in 
Appendix~\ref{stophat}, which describes the evolution of a single 
isolated void.  

Although the image of a large scale matter distribution organized 
by expanding voids is appealing, in its basic form, the description 
essentially involves an extrapolation of single void characteristics 
to an entire random population of strictly distinct and non-interacting 
peers, each of them undisturbed smoothly expanding bubbles. 
This discards one of the most crucial and characteristic aspects of 
cosmic structure formation---that there are no isolated voids, 
nor smoothly unstructured ones. 
Any complete analysis will have to take into account the complications 
which arise from 
\begin{itemize}
 \item{} the substructure present within the primordial volume occupied 
         by the void, and
 \item{} the inhomogeneous matter distribution in its vicinity.
\end{itemize}
The existence of internal void structure is not unexpected. 
The void shown in Figure~\ref{voidcdmcube}, selected from a large 
N-body simulation of cosmic structure formation, shows the existence 
of structure on all scales. The figure shows three successive zoom-ins 
on the inner parts of the void; all exhibit some measure of internal 
structure, although substructure is less pronounced in the emptiest 
inner regions. 

As was mentioned above, all viable cosmological structure formation 
scenarios imply a hierarchical mode of structural growth. 
The formation of any object involves the fusion of all substructure 
present within its realm, including the small-scale objects which had 
condensed out at an earlier stage. 
Underdensities are organized similarly---in the evolution 
of a void we may identify two, intimately related, processes: 
\begin{itemize}
 \item{} a bottom-up assembly, in which a void emerges as a mature and 
         well-defined entity through the fusion and gradual erasure of 
         its internal substructure, and 
 \item{} the interaction of the void with its surroundings, 
         marking its participation in the continuing process of 
         hierarchical structure formation. 
\end{itemize}
Considerable insight into the evolution of voids came from the 
rigorous and insightful study by Dubinski et al.~(1993).  Following 
an analytical study of (isolated) spherically symmetric voids by 
Blumenthal et al.~(1992), they used N-body simulations to study the 
evolution of the void hierarchy from a set of artificial and 
simplified initial conditions, consisting of various levels of 
hierarchically embedded spherical tophat voids. They showed that 
adjacent voids collide, producing thin walls and filaments as the 
matter between them is squeezed. 
Mainly confined to tangential motions, the peculiar velocities 
perpendicular to the void walls are mostly suppressed. 
The subsequent merging of voids is marked by the gradual fading of 
these structures while matter evacuates along the walls and filaments 
towards the enclosing boundary of the ``void merger''. 
The timescale on which the internal substructure of a void is 
erased is approximately the same as that when the void itself 
approached ``nonlinearity'' (Appendix~\ref{stophat} gives a precise 
definition of what is meant by nonlinearity).  
At nonlinearity, smaller-scale voids collide and merge with one 
another, effectively dissolving their separate entities into one 
larger encompassing void. Only a faint and gradually fading imprint of 
their original outline remains as a reminder of the initial internal 
substructure. As this (re)arrangement of structure progresses to ever 
larger scales, the same basic processes repeat.  

\begin{figure*}
\centering
\vskip -5.0cm
\epsfxsize=17.cm
\mbox{\hskip -0.truecm\epsfbox{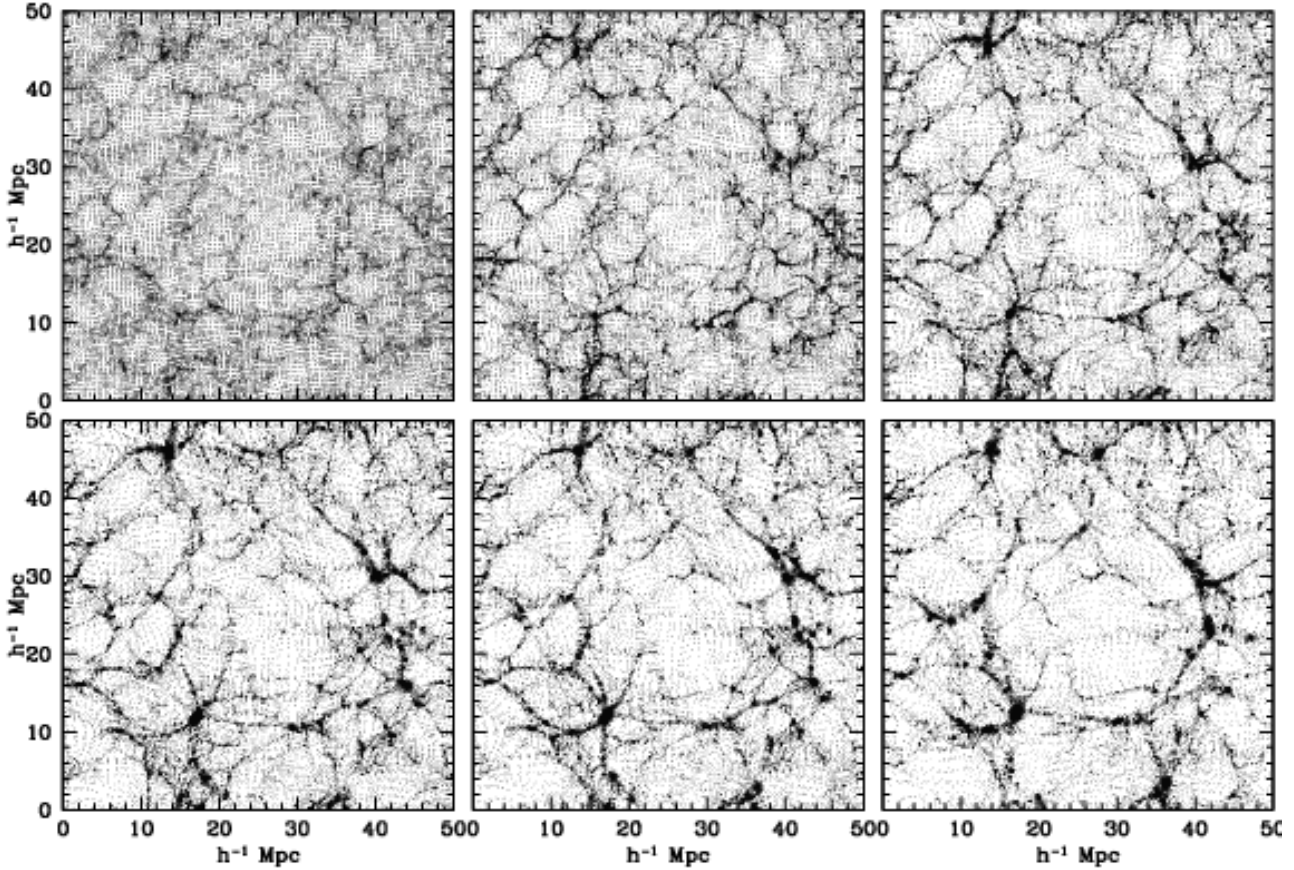}}
\vskip -1.0cm
\caption{Void evolution. 
 Six timesteps in the evolution of a void region in a 128$^3$ 
 particle N-body simulation of structure formation in an SCDM model:  
 top-left to bottom-right shows expansion factors 
 $a_{\rm exp}=0.1, 0.2, 0.3, 0.35, 0.4$ and $0.5$ 
 (the present time has $a_{\rm exp}=1.0$). 
 Initial conditions were defined such that they would focus in on 
 a $3 \sigma (4h^{-1}\hbox{Mpc})$ void, using a constrained random 
 field code (Van de Weygaert \& Bertschinger 1996). 
 The sequence shows the gradual development of a large void of 
 diameter $\approx 25h^{-1}\hbox{Mpc}$ as the complex pattern of 
 smaller voids and structures which had emerged within it at an earlier 
 time, merge with one another. This illustrates one aspect of the 
 evolving void hierarchy:  the {\em void-in-void} process.} 
\label{voidcdm6slc}
\end{figure*}

N-body simulations of voids evolving in more generic cosmological 
circumstances by Van de Weygaert \& Van Kampen~(1993) (also see 
Van de Weygaert~1991) yielded similar results. 
This prompted them to suggest the existence of a natural 
{\it void hierarchy}, in which small-scale voids embedded within a 
pronounced large-scale void gradually fade away.  
An illustration of such a void hierarchy process, within the context 
of the CDM scenario, is shown in Figure~\ref{voidcdm6slc}.  
The major characteristics of an evolving void hierarchy, the 
gradual blending of small-scale voids and structures into a larger 
surrounding underdensity, is clearly visible in the sequence of 
six timesteps. 

However, the artificial arrangement of voids embedded within voids 
represents only one aspect of reality---it misses a crucial component 
of the development of a void hierarchy.  
An evolving void hierarchy not only involves the merging of small 
voids into larger voids, but also {\em the disappearance of small 
voids as they become embedded in larger-scale overdensities.} 
Thus, in contrast to the process of dark halo formation, the 
emerging void hierarchy is ruled by two processes instead of one. 
The main goal of this paper is to incorporate both processes into a 
model of the void hierarchy.  We do this by combining the spherical 
evolution model with the excursion set approach.  When used to describe 
the evolution of overdense clouds, the excursion approach requires 
consideration of a {\it one-barrier} problem, the single barrier 
representing what is required for collapse in the spherical evolution 
model.  We show that the excursion set formulation of the void hierarchy 
requires consideration of {\it two-barriers}:  one barrier is associated 
with the collapse of clouds, and the other with the formation of voids.  
The resulting framework is able to describe realistic settings of 
random density fields in which voids interact with their surroundings. 

This paper is organized as follows.  
Section~\ref{voidevol} discusses important generic properties of 
isolated voids, which grow from depressions in the primordial density 
field, propelled by the perturbed gravitational field.  
The spherical model forms the core of further analytical 
considerations, and is discussed in some detail in Appendix~\ref{stophat}.  
Section~\ref{lss} discusses the generic effects of larger scale 
stucture on the evolution of voids.  {\em Two} crucial processes 
which shape the void hierarchy are described:  
the {\em void-in-void mode} and the {\em void-in-cloud mode}.  
How these processes can be incorporated into the excursion set 
approach using two-barriers is the subject of Section~\ref{excurform}.  
Section~\ref{voiddistr} describes the associated distribution of 
void sizes, which is predicted to have a universal form, and to be 
peaked around a characteristic value.  
One of the results of Section~\ref{voiddistr} is to show that 
peak-based models should be reasonably accurate for the largest 
voids, but, because they account neither for the 
{\em void-in-void mode} nor for the {\em void-in-cloud mode}, they 
predict many more small voids than does the excursion set approach.  
Appendix~\ref{grfs} discusses the {\em ``basic troughs model''}, 
which assumes there is a one-to-one identification between minima 
in the primordial Gaussian density field, with centres of voids in 
the evolved (and nonlinear) matter distribution.  This also serves 
to define notation for  the {\em ``adaptive troughs model''} 
which is described in Section~\ref{ajpeaks}.  

Section~\ref{voidhier} presents various other aspects of the 
hierarchically evolving void population. 
Global parameters, such as the fraction of mass in the cosmos 
contained within void regions, along with the fraction of space 
occupied by voids, are readily derived from the void size distribution. 
In addition, the formalism is applied towards a reconstruction of the 
{\em ancestral} history of a given void, followed by an evaluation of 
the environmental influence on basic void properties. 
We also put forward suggestions towards an analytical treatment of 
the influence of the void environment on the galaxies that may form 
within. Finally, we indicate how an assessment of the evolution of 
dark matter clustering may be predicated on our formalism. 
In Section~\ref{discuss}, we provide an overview of our results and 
seek to embed these in the wider context of the study of hierarchical 
structure formation.  We also comment on how our results for the 
distribution of voids in the dark matter distribution may be related to 
observations of voids in the galaxy distribution.  Although our model 
provides a useful framework, developing a more detailed model is beyond 
the scope of this work.  The results of numerical studies of void 
galaxies in semi-analytic galaxy formation models are described by 
Mathis \& White (2002) and Benson et al. (2003).  

\section{Evolution of isolated voids}\label{voidevol}
The basic features of voids can be understood in terms of the evolution 
of isolated density depressions. 
The net density deficit brings about a sign reversal of the effective 
gravitational force: a void form from a region which induces an effective 
repulsive peculiar gravity.

\begin{figure*}
\centering
\vskip -4.0cm
\epsfxsize=18.5cm
\mbox{\hskip -0.6truecm\epsfbox{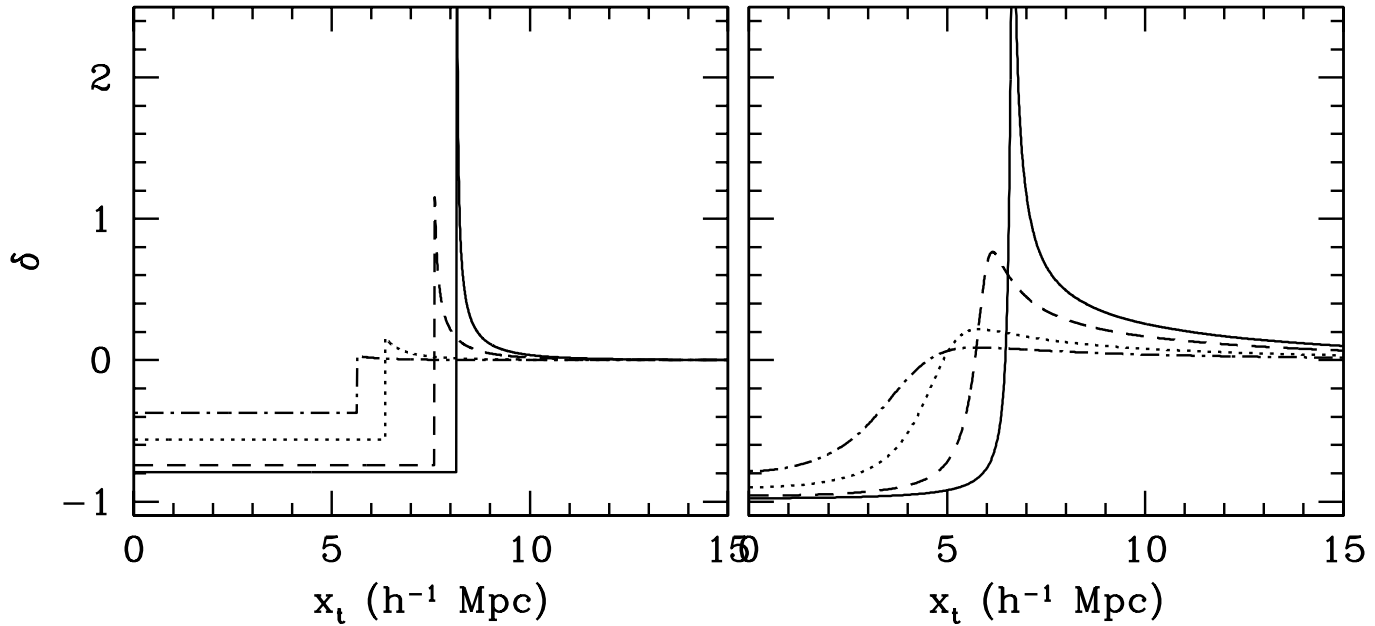}}
\vskip -6.0cm
\caption{Spherical model for the evolution of voids.  
 Left: a pure (uncompensated) tophat void evolving up to the epoch of 
 shell-crossing. Initial (linearly extrapolated) density deficit was 
 $\Delta_{lin,0}=-10.0$, 
 initial (comoving) radius ${\widetilde R}_{i,0}=5.0 h^{-1}\hbox{Mpc}$. 
 Timesteps: $a=0.05, 0.1, 0.2$ and $0.3$. 
 Right: a void with an angular averaged SCDM profile (BBKS, eqn.~7.10). 
 Initial density deficit and characteristic radius are same as for 
 the tophat void (left). The tendency of this void to evolve into a 
 tophat configuration by the time of shell crossing is clear. 
 Shell-crossing, and the formation of an obvious ridge, 
 happens only if the initial profile is sufficiently steep.}
\label{sphmodel}
\end{figure*}

In physical coordinates, overdense regions expand slightly less rapidly 
than the background, reach a maximum size, and then turn around and 
finally collapse to vanishingly small size (this is strictly true only 
in an Einstein de-Sitter or closed Universe). In contrast, underdense 
regions will not turn around:  they undergo simple expansion until 
matter from their interior overtakes the initially outer shells. 
The generic characteristics of these evolutionary paths may be best 
appreciated in terms of the evolution of isolated spherically symmetric 
density perturbations, either overdense or underdense, in an otherwise 
homogeneous and expanding background universe. These spherical models 
provide a key reference for understanding and interpreting more 
complex situations. As a result of the spherical symmetry, 
the problem is essentially one-dimensional, allowing a fully analytic 
treatment and solution, making the model easier to analyze, interpret 
and understand. The spherical model for the evolution of isolated 
voids is discussed in some detail in Appendix~\ref{stophat}.

The most basic and universal properties of evolving spherical voids are:
\begin{itemize}
\item{} {\it Expansion:}
         Voids expand, in contrast to overdense regions, which collapse.
\item{} {\it Evacuation:}
         As they expand, the density within them decreases continuously.
        (To first order, the density decrease is a consequence of the 
         redistribution of mass over the expanding volume. Density 
         decrease from mass lost to the surrounding overdensities is a 
         second order, nonlinear effect occuring only near the edges.) 
\item{} {\it Spherical shape:} 
         Outward expansion makes voids evolve towards a spherical geometry. 
\item{} {\it Tophat density profile:} 
        The effective ``repulsion'' of the matter interior to the 
        void decreases with distance from the center, so the 
        matter distribution evolves into a (reverse) ``tophat''. 
\item{} {\it ``Super-Hubble'' velocity field:} 
        Consistent with its (ultimate) homogeneous interior density 
        distribution, the (peculiar) velocity field in voids has a 
        constant ``Hubble-like'' interior velocity divergence.  
        Thus, voids evolve into genuine ``Super-Hubble Bubbles''.
\item{} {\it Suppressed structure growth:}
        Density inhomogeneities in the interior are suppressed and,  
        as the object begins to resemble an underdense universe, 
        structure formation within it gets frozen-in.  
\item{} {\it Boundary ridge:}
        As matter from the interior accumulates near the 
        boundary, a ridge develops around the void. 
\item{} {\it Shell-crossing:} 
        The transition from a {\it quasi-linear} towards a 
        {\it mature non-linear} stage which occurs as inner shells pass 
        across outer shells. 
\end{itemize}
Figure~\ref{sphmodel} illustrates these features.  
Both panels show the time evolution of the density deficit profile. 
Consider the panel on the left, which illustrates the development of 
an initial (uncompensated) tophat depression (a ``tophat'' void). 
The initial (linear) density deficit of the tophat was set to 
$\Delta_{lin,0}=-10$, and its (comoving) initial radius was 
${\widetilde R}_{i,0}=5h^{-1} \hbox{Mpc}$. 
The evolving density profile bears out the charactertistic tendency 
of voids to expand, with mass streaming out from the interior, and 
hence for the density to continuously decrease in value (and 
approach emptiness, $\delta=-1.0$).  Initially underdense regions 
are just expanding faster than the background and will never collapse 
(in an $\Omega \le 1$ Universe). Notice that this model provides the 
most straightforward illustration of the formation of a ridge. 
Despite the absence of any such feature initially, the void clearly 
builds up a dense and compact bounding ``wall''.

For comparison with the tophat void configuration on the left, 
the panel on the right of Figure~\ref{sphmodel} depicts the 
evolution of a void whose initial configuration is more representative 
of cosmological circumstances. Here, the initial profile is the 
radial-averaged density profile for a trough in a Gaussian random 
field of Cold Dark Matter density fluctuations. 
The analytical expression for this profile was worked out by 
BBKS~\cite{bbks} (eq.~7.10), and the one example we show here concerns 
the radial profile for a density dip with average steepness, i.e. 
$x \equiv -\langle \nabla^2 f \rangle/\sigma_2=-1$. 
The same qualitative aspects of void evolution can be recognized as in 
the case of a pure tophat void: 
 the void expands, 
 empties (to a near-empty configuration $\Delta=-1$ at the centre), 
 and also develops a ridge at its boundary. 
Notice that the void profile evolves into a configuration which 
increasingly resembles that of a ``tophat'' void.  
We will make use of this generic evolution in what follows.  

Looking from the inside out, one sees the interior shells expanding 
outward more rapidly than the outer shells. With a minimum density 
near the void's centre, and density which increases gradually as 
one moves outward, the density deficit $|\Delta(r)|$ of the void 
decreases as a function of radius $r$. The outward directed peculiar 
acceleration is directly proportional to the integrated density 
deficit $\Delta(r,t)$ and therefore decreases with radius: 
inner shells are propelled outward at a higher rate so that the 
interior layers of the void move outward more rapidly. 
The inner matter starts to catch up with the outer shells, leading to 
a steepening of the density profile in the outer realms. 
Meanwhile, over a growing area of the void interior, the density 
distribution is rapidly flattening. This is a direct consequence of 
the outward expansion of the inner void layers: the ``flat'' part of 
the density distribution in the immediate vicinity of the dip gets 
``inflated'' along with the void expansion. 

The features summarized above, which are seen in the idealized 
setting of initially smooth spherically symmetric voids, are also 
seen in more generic, less symmetric cosmological circumstances, when 
substructure is also present.  Figure~\ref{voidcdm6slc} provides one 
illustration of the evolution of more realistic an complex 
underdensities.  N-body simulation studies of objects like this one 
have concluded that the tophat spherical model represent a remarkably 
succesfull desciption of reality (e.g. Dubinski et al.~1993; 
Van de Weygaert \& Van Kampen~1993). 
The evolution towards a spherical top hat, whatever the initial 
configuration, is in stark contrast to how overdensities evolve. 
As a generic overdensity collapses, it contracts along a sequence of 
increasingly anisotropic configurations. 
Contraction leads to a ``deflation'' and accompanying steepening of 
density gradients, while the infall of surrounding structures marks a 
decreasing domain over which the neglect of substructure is 
realistic. 

In summary, it is apparent that the top hat spherical model not only 
provides a rather useful model for the evolution of isolated voids, 
but that it develops into an increasingly accurate representation of 
reality over an increasingly large fraction of the expanding volume 
of the void.

\section{Effect of larger-scale structures}\label{lss}
If we wish to use voids to understand the complex spatial patterns 
in the universe, we need a prescription for identifying the present-day 
cosmic voids, and for describing how voids interact with the large-scale 
structure which surrounds them. 

\subsection{Importance of shell-crossing}
The generic property of ridge formation (e.g. Fig.~\ref{sphmodel}) 
is suggestive, and Blumenthal et al.~\cite{bdglp92} argued that the 
observed voids in the galaxy distribution should be identified with 
primordial underdensities that have only just reached {\it shell-crossing}. 
For a perfectly spherical void with a perfect tophat profile this 
happens exactly when the primordial density depression out of which 
the void developed would have reached a linearly extrapolated 
underdensity $\delta_{\rm v}$.  For such voids, $\delta_{\rm v}$ 
is independent of mass scale:  
 $\delta_{\rm v}=-2.81$ in an $\Omega_0=1$ Universe.  
This threshold value will play an important role in our model of how 
the void hierarchy evolves.  For instance, Dubinski et al.~\cite{ddglp93} 
used this characteristic density to estimate that {\it shell-crossing} 
voids constitute a population of approximately volume-filling domains 
for a substantial range of cosmological structure formation scenarios.
By contrast, overdense primordial perturbations collapse and 
virialize---they {\em shrink} in comoving coordinates. 
The resulting picture is one in which the matter in the Universe 
accumulates in ever smaller collapsing overdensities -- in sheets, 
filaments and clusters -- whose spatial arrangement is dictated by 
the growing underdense expanses.

\begin{figure*}
\centering
\epsfxsize=18.5cm
\epsfbox{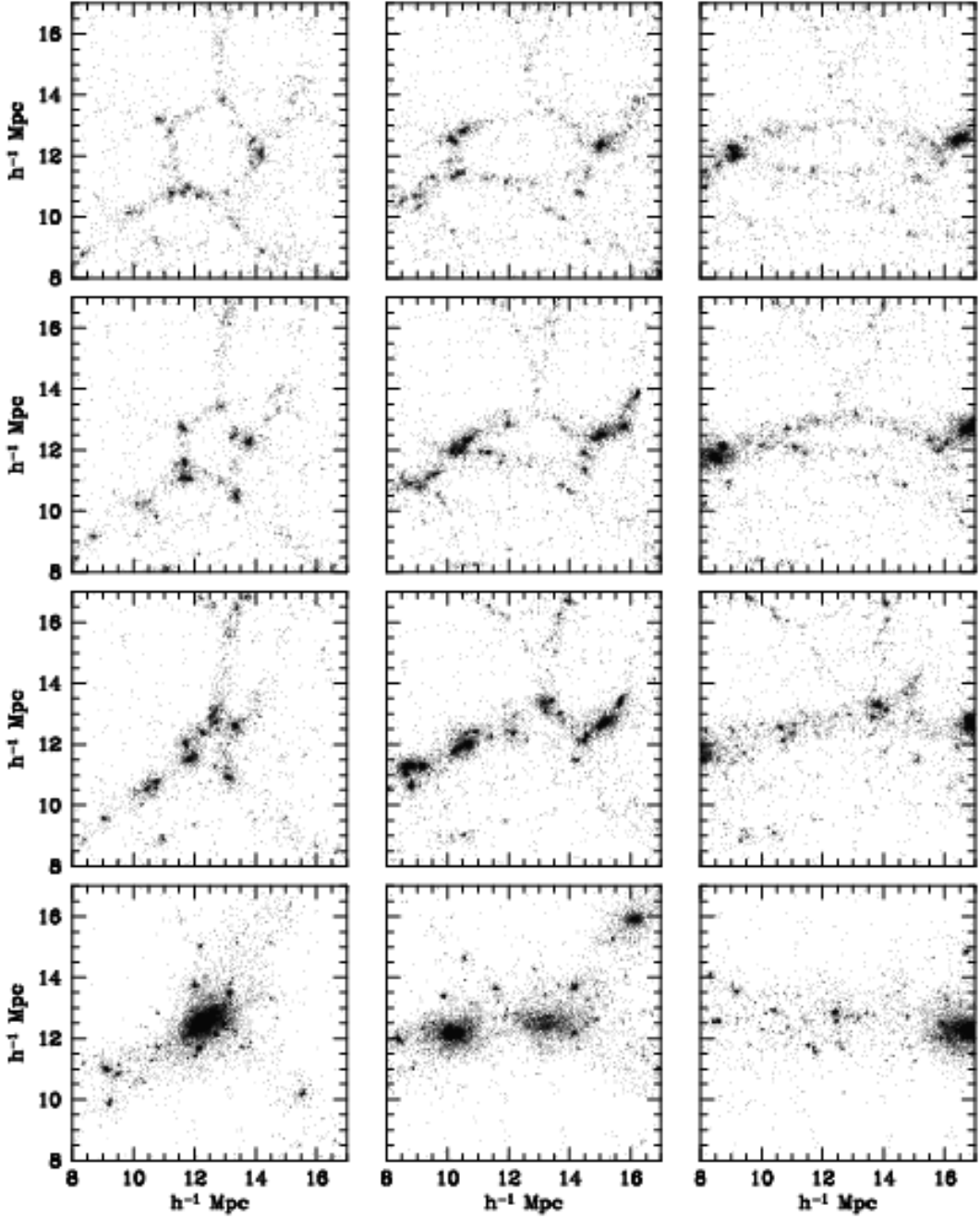}
\vskip -1.0truecm
\caption{Three examples (left to right) of the {\em void-in-cloud} process 
in action in numerical simulations of structure formation in a SCDM scenario 
($\Omega_0=1.0$, $h_0=0.5$). Top to bottom panels show the evolution 
of the particle distribution in comoving coordinates from early to late 
times (resp. $a=0.3, 0.4, 0.5$ and $1.0$, the current epoch). The initially 
underdense regions are crushed by the collapse of the overdense regions which 
surround them. The void in the first row of 4 panels shows a nearly 
spherical collapse sequence. The other two rows involve configurations 
involving more anisotropic surrounding matter distributions (and force 
fields).}
\label{voidcollnbody}
\end{figure*}

\subsection{Void Sociology}
Two effects will seriously affect the number of small voids within 
a generic field of density perturbations. 
Both relate to the hierarchical embedding of a density depression within 
the larger scale environment. 

First, consider a small region which was less dense than the critical 
$\delta_{\rm v}$.  
It may be that this region, which we would like identify as a void today, 
was embedded in a significantly larger underdense region which was also 
less dense than the critical density.  Therefore, we would also like to 
identify the larger region as a large void today.  Since many small 
voids may coexist within one larger void, we must not count all of the 
smaller voids as distinct objects, lest we overestimate the number of 
small voids, and the total volume fraction in voids. 
We will call this the {\it void-in-void} problem. It is analogous 
to the well-known {\em cloud-in-cloud} problem associated with the 
using the number density of initially overdense peaks to estimate 
the number of dense virialized clusters. 

A {\em second} effect is responsible for a radical dissimilarity 
between void and halo populations:  If a small scale minimum is embedded 
in a sufficiently high large scale maximum, then the collapse of the 
larger surrounding region will eventually squeeze the underdense 
region it surrounds; the small-scale void will vanish when the region 
around it has collapsed completely. If the void within the contracting 
overdensity has been squeezed to vanishingly small size it should no 
longer be counted as a void. Figure~\ref{voidcollnbody} shows three 
examples of this process, each identified from a large (SCDM) N-body 
simulation. To account for the impact of voids disappearing when 
embedded in collapsing regions, we must also deal with the 
{\it void-in-cloud} problem. 

Virialized halos within voids are not likely to be torn apart as the 
void expands around them.  Thus, the {\em cloud-in-void} phenomenon 
is irrelevant for dark halo formation. 
The asymmetry between the {\em void-in-cloud} and {\em cloud-in-void} 
processes effects a symmetry breaking between the emerging halo and 
void populations:  although they evolve out of the same symmetric 
Gaussian initial conditions, we argue that over- and underdensities 
are expected to evolve naturally into agglomerations with rather  
different characteristics.  

\section{Excursion Formalism}\label{excurform}
In its simplest and most transparent formulation the excursion set 
formalism refers to the collapse of perfectly spherical 
overdensities, so this is the case which we will describe first.

\begin{figure}
\centering
\vskip -1.7truecm
\epsfxsize=9.8cm
\mbox{\hskip -0.85truecm\epsfbox{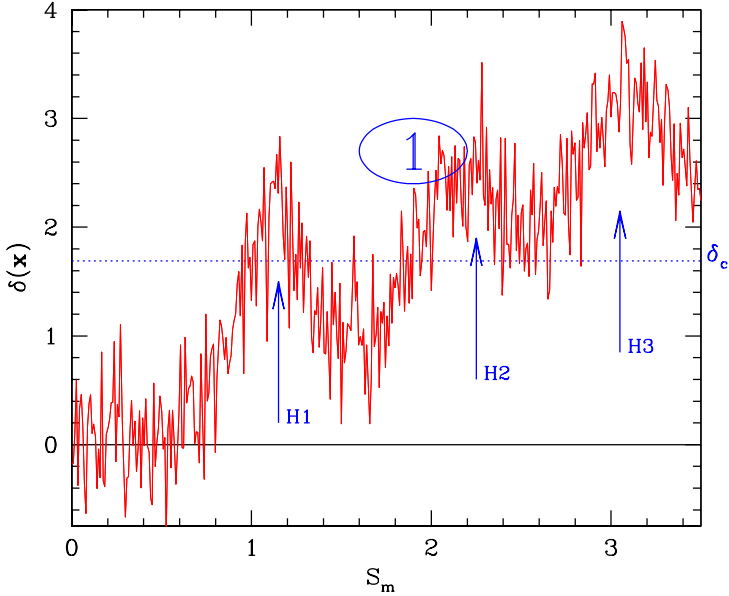}}
\vskip -0.0truecm
\caption{Excursion Set Formalism, illustrated for the formation 
of a halo. Random walk exhibited by the average overdensity $\delta$ 
centred on a randomly chosen position in a Gaussian random field, as a 
function of smoothing scale, parametrized by $S_m$ (large volume are 
on the left, small volumes on the right).  
Dashed horizontal line indicates the collapse barrier $\delta_{\rm c}$.  
The largest scale (smallest value of $S$) on which $\delta(S)$ exceeds 
$\delta_{\rm c}$ is an estimate of the mass of the halo which will 
form around that region.  }
\label{excurhalo}
\end{figure}

\subsection{Excursion set model of clusters}\label{walks}
The jagged line in Fig.~\ref{excurhalo} represents the overdensity 
centred on a randomly chosen position in the initial Gaussian random 
field, as a function of the scale on which the overdensity was 
computed.  
The height of the walk $\delta_0(S)$ is the linear theory overdensity 
relative to the density of the background universe. The spatial scale 
is parametrized by its variance $S$ (defined in equation~\ref{sigmaj}).  
In hierarchical models, $S$ decreases with increasing scale, so the 
largest spatial scales are on the left, and $\delta(S)\to 0$ as $S\to 0$.  
Because the initial fluctuations are small, the mass contained within 
the smoothing filter is $m\propto [1+(D_i/D_0)\delta_0]\,R^3$, 
where $D_i$ denotes linear theory growth factor at the initial time.  
Since $D_i/D_0\ll 1$, $m\propto R^3$:  the mass is proportional to 
the initial comoving scale cubed.  

In the spherical collapse model, all regions with linear theory 
densities greater than $\delta_{\rm c}$ can have formed bound 
virialized objects, and this critical overdensity is independent of 
mass scale.  This constant value is shown as the dashed line in 
same height at all $S_m$, where we have used the subscript $m$ to 
denote the fact that mass and initial scale are interchangeable.  . 

The excursion set formalism supposes that no mass can escape from a 
region which collapses. If $\delta_0=\delta_{\rm c}$ on scale $R$, 
then all the mass contained within $R$ is included in the collapsed 
object, even if $\delta_0 < \delta_{\rm c}$ for all $r<R$. Thus, if 
the random walk height $\delta_0$ exceeds the value $\delta_{\rm c}$ 
after having travelled distance $S(R)$ it represents a collapsed object 
of mass $m\propto R^3$.  
A walk may cross the barrier $\delta_{\rm c}$ at many different values 
of $S(R)$. Each crossing corresponds to a different smoothing scale and, 
because $m\propto R^3$, contains a different amount of mass. However, of the 
various crossings of the barrier $\delta_{\rm c}$ the first crossing, at the 
smallest value of $S(R)$ for which $\delta_0\ge \delta_{\rm c}$, is special 
since it is this scale which is associated with the most mass.  
The crossings at smaller scales correspond to condensations of a 
smaller mass, which have been incorporated in the larger encompassing 
mass concentration.  

In its simplest form, the excursion model for the distribution of 
masses of virialized objects equates the distribution of distances 
$S(R)$ which one-dimensional Brownian motion random walks, 
originating at the origin, travel before they first cross a barrier 
of constant height $\delta_{\rm c}$, with the fraction of mass which 
is bound up in objects of mass $m(R)$. 
The further a given walk travels before crossing the barrier, the 
smaller the mass of the object with which it is associated 
(Bond et al. 1991).

\begin{figure*}
\centering
\vskip 0.0cm
\epsfxsize=19.cm
\mbox{\hskip 0.5truecm\epsfbox{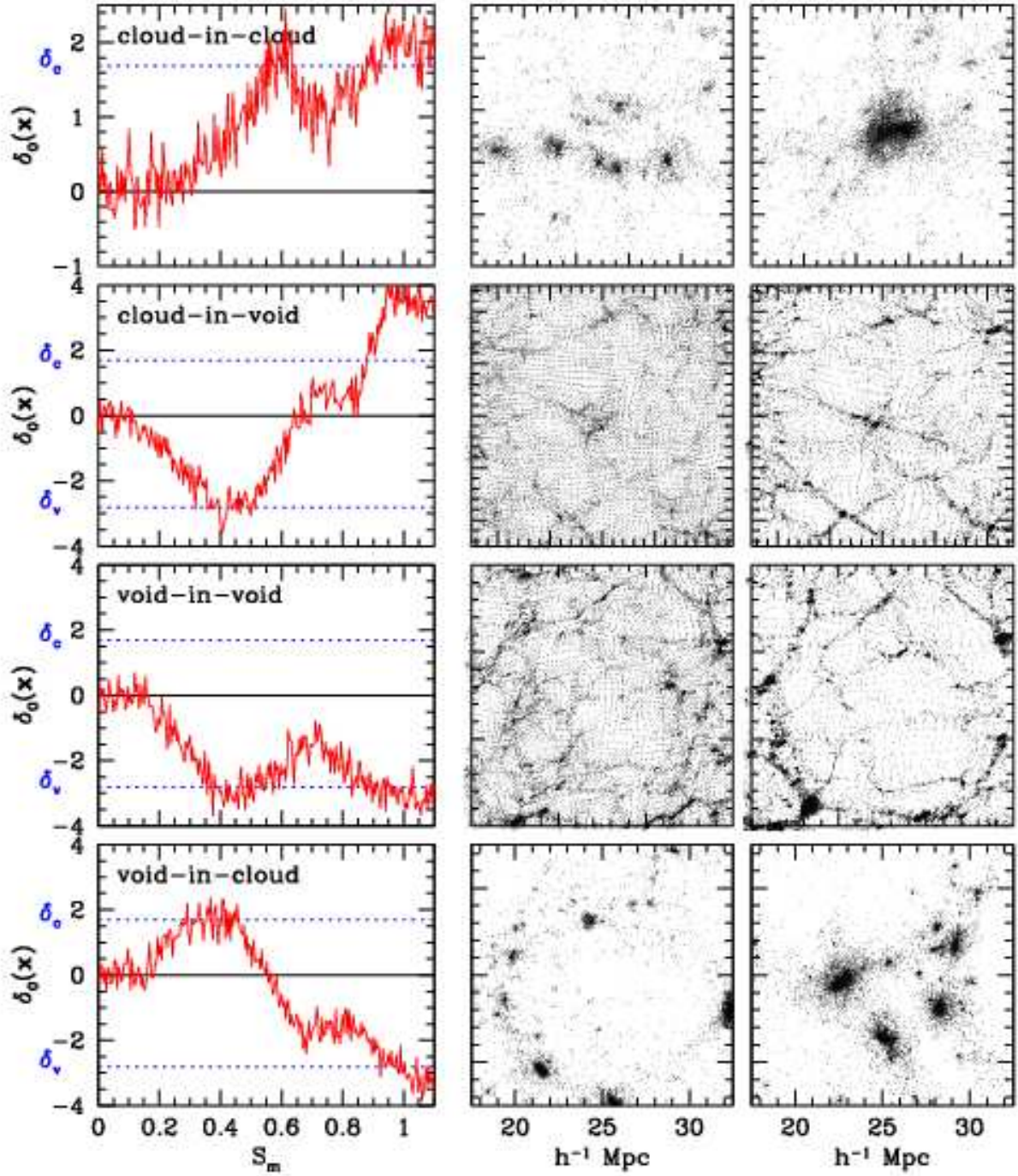}}
\vskip 0.0cm
\caption{Four mode (extended) excursion set formalism. 
Each row illustrates one of the four basic modes of hierarchical 
clustering: the {\em cloud-in-cloud} process, 
{\em cloud-in-void} process, {\em void-in-void} process and 
{\em void-in-cloud} process (from top to bottom). 
Each mode is illustrated using three frames. 
Leftmost panels show `random walks': the local density 
perturbation $\delta_0({\bf x})$ as a function of (mass) 
resolution scale $S_m$ (cf. Fig.~\ref{excurhalo}) at an early 
time in an N-body simulation of cosmic structure formation. 
In each graph, the dashed horizontal lines indicate the 
{\em collapse barrier} $\delta_{\rm c}$ and the shell-crossing 
{\em void barrier} $\delta_{\rm v}$. 
The two frames on the right show how the associated particle 
distribution evolves. 
Whereas halos within voids may be observable (second row depicts a 
halo within a larger void), voids within collapsed halos are not 
(last row depicts a small void which will be squeezed to small size 
as the surrounding halo collapses). It is this fact which makes the 
calculation of void sizes qualitatively different from that usually 
used to estimate the mass function of collapsed halos.}
\label{excurcensus}
\end{figure*}

\subsection{Excursion set model of voids}\label{vexcursion}
In our discussion above of the halo mass function, 
we considered the {\em cloud-in-cloud} problem, and argued that the 
only cloud which should be counted was the largest possible one.  
To study voids in the excursion set approach one must first specify 
the boundary shape associated with the emergence of a void. This can 
be done if we know the critical underdensity $\delta_{\rm v}$ which 
defines a void, and in what follows we will use the epoch of 
shell-crossing, estimated using the spherical evolution model, to 
specify $\delta_{\rm v}$.   Thus, $\delta_{\rm v}=-2.81$, independent 
of smoothing scale (as was $\delta_{\rm c}$).  

One might have thought that whereas clusters form from overdensities, 
voids form from underdensities, so the distribution of voids can be 
estimated analogously to how one estimates the distribution of 
clusters --- one simply replaces the barrier $\delta_{\rm c}$ with 
one at $\delta_{\rm v}$, and then studies the distribution of first 
crossings of $\delta_{\rm v}$.  
Thus, if the random walk $\delta_0$ first drops below the value 
$\delta_{\rm v}$ after having travelled distance $S(R)$ it represents 
a void of mass $m\propto R^3$ and physical size ${\cal R} \approx 1.7R$. 

However, we have seen that we must be more careful; in addition 
to avoiding the double counting associated with the void-in-void process, 
we must also account for the void-in-cloud process.  The strength of 
the excursion set formulation is that is shows clearly how to do this.  
Figure~\ref{excurcensus} illustrates the argument.  
There are four sets of panels.  The left-most of each set shows the 
random walk associated with the initial particle distribution.  
The two other panels show how the same particles are distributed 
at two later times.  
The first set illustrates the cloud-in-cloud process.  
The mass which makes up the final object (far right) is given by 
finding that scale within which the linear theory variance has 
value $S=0.55$.  This mass came from the mergers of the smaller 
clumps, which themselves had formed at earlier times (centre panel).  
If we were to center the random walk path on one of these small 
clumps, it would cross the higher barrier 
$\delta_{\rm c}/D(t)>\delta_{\rm c}$ at $S>0.55$, the value of 
$D(t)$ representing the linear theory growth factor at the earlier 
time $t$.

The second series of panels shows the cloud-in-void process.  
Here, a low mass clump ($S>0.85$) virializes at some early time.  
This clump is embedded in a region which is destined to become a 
void.  The larger void region around it actually becomes a bona-fide 
void only at the present time, at which time it contains significantly 
more mass ($S=0.4$) than is contained in the low mass clump at its 
centre.  Notice that the cloud within the void was not destroyed 
by the formation of the void; indeed, its mass increased slightly 
from $S>0.85$ to $S\sim 0.85$.  Such a random walk is a bona-fide 
representative of $S\sim 0.85$ halos; for estimating halo abundances, 
the presence of a barrier at $\delta_{\rm v}$ is irrelevant.  
On the other hand, walks such as this one allow us to make some 
important inferences about the properties of void-galaxies, which we 
will discuss shortly.  

The third series of panels shows the formation of a large void by 
the mergers of smaller voids:  the void-in-void process.  
The associated random walk looks very much the inverse of that 
for the cloud-in-cloud process associated with halo mergers.  
The associated random walk shows that the void contains more 
mass at the present time ($S\sim 0.4$) than it did in the past ($S>0.4$); 
it is a bona-fide representative of voids of mass $S\sim 0.4$.  
A random walk path centered on one of these mass elements which make 
up the filaments within the large void would resemble the 
cloud-in-void walk shown in the second series of panels.  
[Note that the height of the barrier associated with voids which 
are identified at cosmic epoch $t$ scales similarly to the barrier 
height associated with halo formation:  
 $\delta_{\rm v}(t)\equiv \delta_{\rm v}/D(t)$.]

Finally, the fourth series of panels illustrates the {\em void-in-cloud} 
process.  The particle distribution shows a relatively large void at 
the early time being squeezed to a much smaller size as the ring of 
objects around it collapses.  A simple inversion of the cloud-in-void 
argument would have tempted one to count the void as a relatively 
large object containing mass $S\sim 1$.  That this is incorrect can be 
seen from the fact that, if we were counting halos, we would have 
counted this as a cloud containing significantly more mass ($S\sim 0.3$), 
and it does not make sense for a massive virialized halo to host a 
large void inside.  

Thus, the excursion set model for voids which we will develop 
below is as follows:  
If a walk first crosses $\delta_{\rm c}$ and then crosses 
$\delta_{\rm v}$ on a smaller scale, then the smaller void is 
contained within a larger collapsed region. 
Since the larger region has collapsed, the smaller void within it 
no longer exists, so it should not be counted. 
The only bona-fide voids are those associated with walks which 
cross $\delta_{\rm v}$ without first crossing $\delta_{\rm c}$.  
{\em The problem of estimating the fraction of mass in voids reduces 
to estimating the fraction of random walks which first crossed 
$\delta_{\rm v}$ at $S$, and which did not cross $\delta_{\rm c}$ at 
any $S'<S$}.  Thus, a description of the void hierarchy requires 
solution of a {\em two-barrier} problem. 

Clearly, the model predictions will depend on $\delta_{\rm c}$ and 
$\delta_{\rm v}$.  If we use the spherical tophat model summarized 
in Appendix~\ref{stophat} to set these values, then it seems reasonable 
to set $\delta_{\rm v}=-2.81$.  But how we account for the 
void-in-cloud problem is somewhat more subtle. 
Suppose we choose $\delta_{\rm c}=1.686$, the value associated 
with complete collapse.  In effect, this allows a void to have the 
maximum possible size it can have, given its underdensity, unless 
it is within a fully collapsed halo, in which case it has zero 
size.  Presumably, if it is within a collapsing region which has 
not yet collapsed completely (as in the bottom-right panel of 
Figure~\ref{excurcensus}), then its size is intermediate between 
the size one would have estimated from the isolated spherical 
evolution model, and zero.  Thus, only excluding voids in regions 
which have collapsed completely almost certainly overestimates the 
typical void size (furthermore, we are ignoring the thickness of 
the ridge around each void).  
Another natural choice is $\delta_{\rm ta}=1.06$; this ignores all 
voids that are within regions which are beginning to turnaround, 
even though they may still have non-neglibigle sizes, and so 
underestimates the abundance of large voids.  Accounting more 
carefully for the effect of the void-in-cloud problem is the 
subject of ongoing work.  

In summary, what distinguishes voids from collapsed objects is the 
following:  Whereas it may be possible to have a cluster within a 
void it does not make physical sense to have a void within a cluster. 
The excursion set formulation allows one to account for this.

\section{Universal Void Size Distribution}\label{voiddistr}
Let ${\cal F}(S,\delta_{\rm v},\delta_{\rm c})$ denote the fraction 
of walks which first cross $\delta_{\rm v}$ at $S$, and which do not 
cross $\delta_{\rm c}$ until after they have crossed $\delta_{\rm v}$
(i.e., if they cross $\delta_{\rm c}$, they do so at $s\ge S$).  
Then ${\cal F}(S,\delta_{\rm v},\delta_{\rm c})$ is the distribution 
of first crossings of the type associated with voids. 
Appendix~\ref{fs2bar} shows that this first crossing distribution 
is given by 
\begin{eqnarray}
 S\,{\cal F}(S,\delta_{\rm v},\delta_{\rm c}) &=& 
           \sum_{j=1}^\infty {j^2\pi^2 {\cal D}^2\over\delta_{\rm v}^2/S}\,
           {\sin(j\pi {\cal D})\over j\pi} ,
\nonumber\\
 && \qquad \times 
    \exp\left(-{j^2\pi^2 {\cal D}^2\over 2\,\delta_{\rm v}^2/S}\right),
 \label{vfvoid}
\end{eqnarray}
where 
\begin{equation}
{\cal D} \equiv {|\delta_{\rm v}|\over (\delta_{\rm c}+|\delta_{\rm v}|)}\,.  
\end{equation}
In an Einstein de-Sitter universe, $\delta_{\rm c}$, $\delta_{\rm v}$ 
and $\sigma(m)$ all have the same time dependence, 
so equation~(\ref{vfvoid}) evolves self-similarly.  
In more general world-models the time dependences only slightly 
different, so the approximation of self-similar evolution should be 
quite accurate.  

The quantity ${\cal D}$ is the ``{\em void-and-cloud parameter}'';  
it parameterizes the impact of halo evolution on the evolving 
population of voids.  To see why, notice that the likelihood of 
smaller voids being crushed through the {\em void-in-cloud} process 
decreases as the relative value of the collapse barrier $\delta_{\rm c}$ 
with respect to the void barrier $\delta_{\rm v}$ becomes larger. 

This is also consistent with the fact that 
\begin{equation}
 \int {\cal F}(S,\delta_{\rm v},\delta_{\rm c}) = 1-{\cal D} 
      = {\delta_{\rm c}\over\delta_{\rm c}+|\delta_{\rm v}|},
 \label{massfrac}
\end{equation}
(e.g., equation~\ref{lt0}) represents the {\em mass fraction in voids}.  
Thus, if ${\cal D}$ is small, voids account for nearly all the mass. 
On the other hand, for any noticeable impact of the void-in-cloud 
process the mass fraction in voids, $1-{\cal D}$, will be less than 
unity. The more important the void-in-cloud process is, the smaller 
the mass fraction in voids will be, as more voids are squeezed to 
vanishingly small size. 

Relation~(\ref{massvol}) suggests that the volume fraction in voids is 
$1.7^3\,(1-{\cal D})$.  For $\delta_{\rm v}=-2.81$ and 
$\delta_{\rm c}=1.686$, this ratio is larger than unity, indicating 
that the voids fill the universe.   
(The volume fraction in voids is also larger than unity if we set 
$\delta_{\rm c}=1.06$ instead.)  Thus, we have a model in which 
{\em about one third of the mass of the universe is associated with 
voids which occupy most of the volume}.  The remaining seventy percent 
of the mass is in between the voids, and occupies negligible volume.  

Although the sum in equation~(\ref{vfvoid}) converges reasonably 
rapidly, it is not so easy to see what shape it implies.  
We have found that equation~(\ref{vfvoid}) is quite well 
approximated by 
\begin{equation}
 \nu f(\nu) \approx \sqrt{\nu\over 2\pi}\,
 \exp\left(-{\nu\over 2}\right)\,
 \exp\left(-{|\delta_{\rm v}|\over\delta_{\rm c}}\,
 {{\cal D}^2\over 4\nu}-2{{\cal D}^4\over\nu^2} \right) ,
 \label{vfvapprox}
\end{equation}
where we have set 
\begin{equation}
 \nu\,\equiv\,\delta_{\rm v}^2/S \equiv\,\delta_{\rm v}^2/\sigma^2(m)\,,
\end{equation}
and $\nu f(\nu)\,{\rm d}\nu/\nu = S{\cal F}(S)\,{\rm d}S/S$. 
(This expression is accurate for values of 
$\delta_{\rm c}/|\delta_{\rm v}|\ge 1/4$ or so.)   
Expression~(\ref{vfvapprox}) shows clearly that $f(\nu)$ cuts-off 
sharply at both small and large values of $\nu$. In other words, the 
distribution of void masses is reasonably well peaked about $\nu \approx 1$, 
corresponding to a characteristic mass of order $\sigma_0(m)\approx 
|\delta_{\rm v}|$.  

When $\delta_{\rm c}\gg |\delta_{\rm v}|$, then ${\cal D}\to 0$, and the 
second exponential tends to unity. In this limit, the two-barrier 
distribution reduces to that associated with a single barrier at 
$\delta_{\rm v}$. This shows explicitly that when the void-in-cloud 
process is unimportant (${\cal D}\to 0$), then the abundance of voids 
is given by accounting correctly for the void-in-void process.  

\begin{figure}
\vskip -1.4truecm
\epsfxsize=10.8cm
\mbox{\hskip -0.8truecm\epsfbox{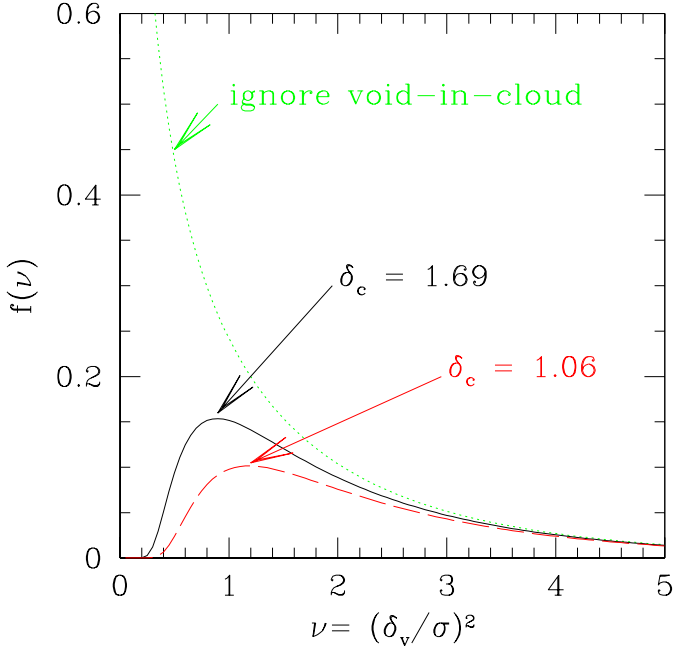}}
\vskip -0.7truecm
 \caption{Scaled distribution of void masses/sizes: 
 voids which enclose large masses have  large values of $\nu$.  
 Curves show equation~(\ref{vfvoid}) with $\delta_{\rm v}=-2.81$ and 
 two choices of $\delta_{\rm c}$ as labeled.  These choices are 
 motivated by the spherical collapse model, and result in a distribution 
 which is well peaked about a characteristic value.  
 Ignoring the void-in-cloud process altogether is equivalent to 
 setting $\delta_{\rm c}\to\infty$. 
 Although decreasing $\delta_{\rm c}/|\delta_{\rm v}|$ decreases 
 the abundance of low mass halos, the abundances of voids which 
 enclose the most mass are not sensitive to the value of $\delta_{\rm c}$.}
\label{fvlin}
\end{figure}

Figure~\ref{fvlin} illustrates the resulting void size distributions. 
Notice that the mass fraction in 
small voids depends strongly on $\delta_{\rm c}$ (the divergence at 
low $\nu$ associated with the void-in-void solution is removed as 
$\delta_{\rm c}$ increases), whereas the mass fraction enclosed by 
the largest voids depends only on $\delta_{\rm v}$.  
This is primarily a consequence of the fact that large underdensities 
embedded in a larger region of average density are rare, so such 
regions embedded in large overdensities are rarer still.  
Since there are essentially no large-scale underdensities 
embedded in larger scale overdensities, on scales where 
$\sigma\ll (\delta_{\rm c}+|\delta_{\rm v}|)$, 
the value of $\delta_{\rm c}$ is irrelevant.  
Thus, the distribution of large voids is almost exclusively determined 
by $\delta_{\rm v}$.  We will return to this shortly.  

The number density $n(m)$ of voids which contain mass $m$ is obtained 
by inserting expression~(\ref{vfvoid}) in the relation 
\begin{equation}
 {m^2\,n_{\rm v}(m)\over\bar\rho} = 
 S\,{\cal F}(S,\delta_{\rm v},\delta_{\rm c}) \,
 {{\rm d\ ln}\,S\over {\rm d\ ln}\,m}.
\label{fmassvoid}
\end{equation}
To illustrate what our two-barrier model implies for void sizes, 
we must convert the expression above for the fraction of mass in 
voids to a void-size distribution. The simplest approximation, 
motivated by the spherical tophat void model, sets the comoving 
volume $v$ of the void equal to 
\begin{equation}
 v = (m/\bar\rho) \times 1.7^3\,.
 \label{massvol}
\end{equation} 
Since all the time dependence enters via  
$\nu=\delta^2_{\rm v}(z)/\sigma^2(m)$, 
the distribution of void sizes evolves self-similarly.  
Simple changes of variables relate the void volume or mass functions 
to the barrier crossing distribution:  
 $f(v){\rm d}v = f(m){\rm d}m = f(\nu){\rm d}\nu$.  

\begin{figure*}
\vskip -2.5truecm
\epsfxsize=19.2cm
\mbox{\hskip -0.3truecm\epsfbox{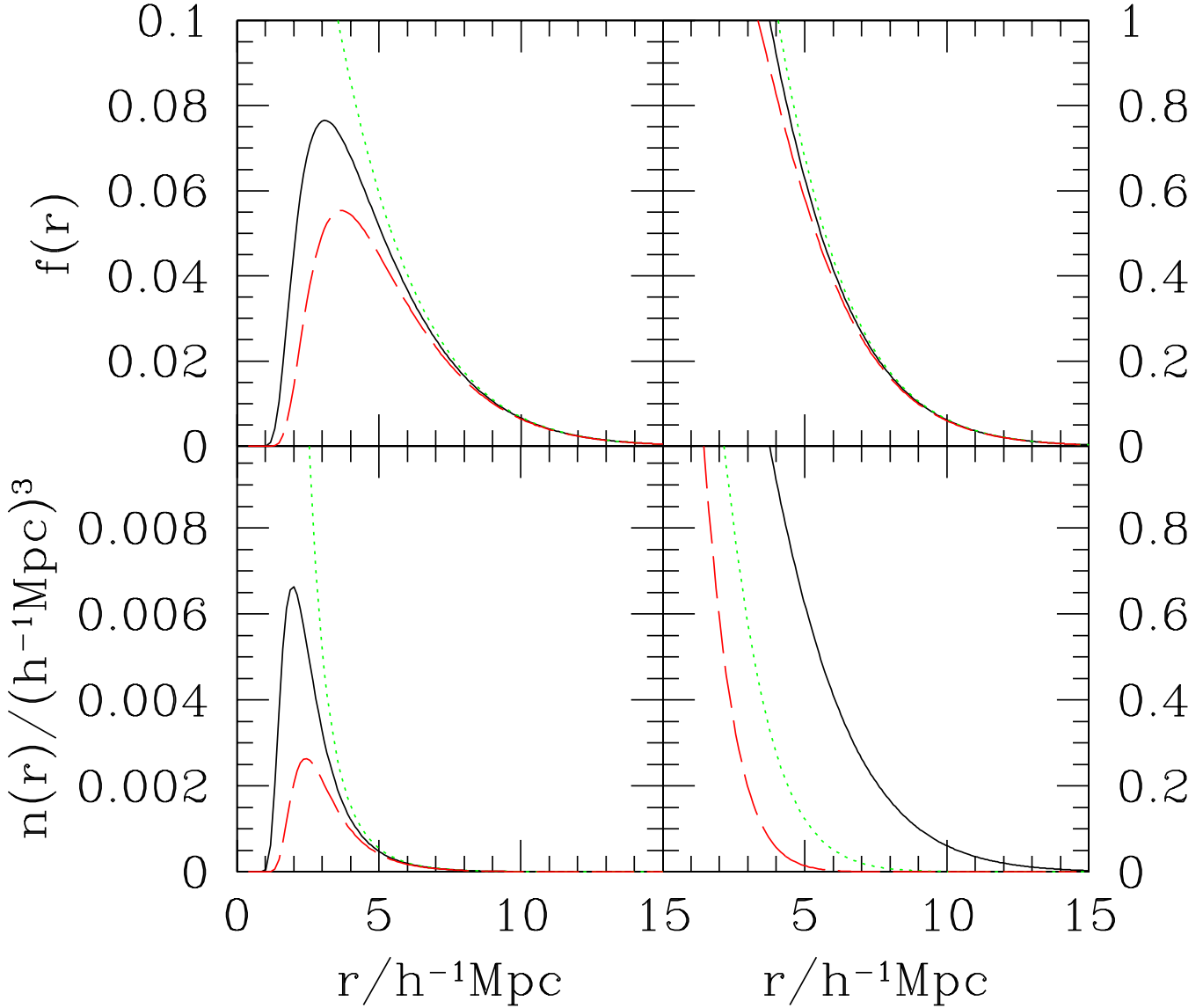}}
\vskip -1.25truecm
 \caption{Distribution of void radii predicted by 
 equation~(\ref{vfvapprox}), in an Einstein de-Sitter model with 
 $P(k)\propto k^{-1.5}$, normalized to $\sigma_8=0.9$ at $z=0$.  
 Top left panel shows the mass fraction in voids of radius $r$.  
 Bottom left panel shows the number density of voids of radius $r$.  
 Note that the void-size distribution is well peaked about a 
 characteristic size provided one accounts for the void-in-cloud 
 process.  Top right panel shows the cumulative distribution of the 
 void volume fraction. 
 Dashed and solid curves in the top panels and bottom left panel 
 show the two natural choices for the importance of the void-in-cloud 
 process discussed in the text:  $\delta_{\rm c}=1.06$ and 1.686, 
 with $\delta_{\rm v}=-2.81$.  Dotted curve shows the result of 
 ignoring the {\em void-in-cloud} process entirely.
 Clearly, the number of small voids decreases as the ratio of 
 $\delta_{\rm c}/|\delta_{\rm v}|$ decreases. 
 Bottom right panel shows the evolution of the cumulative void volume 
 fraction distribution. The three curves in this panel are for 
 $\delta_{\rm c}=1.686(1+z)$, where $z=0$ (solid), 0.5 (dotted) 
 and~1 (dashed).}
\label{frvoid}
\end{figure*}

As a specific illustration of what our model implies, 
Fig.~\ref{frvoid} shows the distribution of void sizes in a model 
where the initial power-spectrum was $P(k)\propto k^n$ with $n=-1.5$, 
normalized so that the rms fluctuations in a tophat sphere of 
radius unity was $\sigma_8=0.9$ at $z=0$. The top left panel shows 
the mass fraction in voids of radius $r$, and the bottom left panel 
shows the number density of such voids.  The three curves in each panel 
show equation~(\ref{vfvapprox}) with $\delta_{\rm c}=1.06$, 1.686 and 
$\infty$, and we have set $\delta_{\rm v}=-2.81$ in all cases.  
Notice how the abundance of small voids decreases dramatically as the 
ratio $\delta_{\rm c}/|\delta_{\rm v}|$ decreases. By contrast, the 
abundance of large-scale voids is largely insensitive to this ratio 
(also see Fig.~\ref{fvlin}). 

We can make a rough estimate of the scale of the peak by computing 
that $\nu$ at which equation~(\ref{vfvapprox}) is maximized.  
This requires solution of a cubic, and gives $\nu_{\rm max}$ 
decreasing as ${\cal D}$ decreases. For the range of $0.6\le {\cal D}\le 0.75$ 
of interest, it is usually close to unity: $\nu_{\rm max} \approx 1 $.
To estimate the typical void size we will therefore simply use the 
approximate value of $\sigma\sim |\delta_{\rm v}|$. 

For a power spectrum approximated by a power-law of slope $n$, the 
initial comoving size $R_i$ of a region which is identifed 
as a void is, 
\begin{equation}
 {|\delta_{\rm v}|\over\sigma_8} = 
 \left({8\over R_{\rm i}}\right)^{(n+3)/2} \quad\Rightarrow\quad 
 R_{\rm i} = 8\,\left({\sigma_8\over |\delta_{\rm v}|}\right)^{2/(n+3)}\,
\end{equation}
with $\sigma_8$ denoting the rms fluctuation on scales of $8h^{-1}$Mpc 
(currently favoured $\Lambda$CDM models have $\sigma_8\approx 0.9$). 
This means that the final size $r_v$ of the void is 
\begin{equation}
 {r_{v}\over h^{-1} {\rm Mpc}}\sim 1.7\times {8\over 3^{2/(3+n)}}\,
 \left({\sigma_8\over 0.9}{2.7\over |\delta_{\rm v}|}\right)^{2/(3+n)}.
\end{equation}
\noindent A reasonable approximation to CDM spectra on Megaparsec 
scales is obtained by setting $n=-1.5$.  In this case, the typical 
void radius is $\sim 3h^{-1}$Mpc.  Since the correlation length is of 
order $8h^{-1}$Mpc, this makes the typical void diameter of order the 
correlation length.  

The top right panel shows the cumulative distribution of the volume 
fraction for the three choices of $\delta_{\rm c}$.  In all three 
cases, voids with radii greater than $5h^{-1}$ account for about sixty 
percent of the volume.  This suggests that, for sufficiently large 
voids, the details of the {\em void-in-cloud} process are not important.  
It is easy to see why:  a typical cluster forms from a region which had 
comoving radius 
 $R_i \sim 8\,(\sigma_8/\delta_{\rm c})^{2/(3+n)} \sim 3.5h^{-1}$Mpc.  
Since few collapsing regions are larger than this, voids which are 
initially larger than this are extremely unlikely to have been squeezed 
out of existence.  

Finally we turn to an estimate of how the volume fraction in voids 
evolves in this model.  
Since $\sigma_8(z)=\sigma_8/(1+z)$, the typical comoving size of 
voids is expected to be smaller at higher redshifts, by a factor 
of $(1+z)^{-2/(3+n)}$.  The bottom panel shows the cumulative 
distribution at redshifts zero, one-half and unity 
(solid, dotted and dashed curves) where we have approximated 
$\delta_{\rm c}(z)\,=\,1.686(1+z)$ and $\delta_{\rm v}(z) = -2.81(1+z)$.

\subsection{Alternative Models}\label{ajpeaks}
To better appreciate the ramifications of the 
{\it two-barrier excursion set} model, it is instructive to explore 
alternative descriptions. This section discusses two models which 
follow from associating present-day voids with sufficiently underdense 
troughs in the initial fluctuation field.  

\subsubsection{The Basic Troughs Model}
The most straightforward model of the void distribution is to suppose that 
voids are associated with minima in the initial density field. The simplest 
approximation to the number density of voids comes from smoothing the initial 
density fluctuation field with a filter of scale $R$, and then counting the 
number of minima of depth $\delta_{\rm v}$ in the smoothed field. 
If one assumes that all the initial minima survive to the present 
time, then the number density of minima gives the number density of 
voids. BBKS~\cite{bbks} show that the density of minima of depth 
\begin{equation}
 \nu\ =\ {\displaystyle \delta^2_{\rm v} \over \displaystyle \sigma^2_0(m)}\,, 
\end{equation}
in a Gaussian random field is 
\begin{equation}
 n_{\rm v}(\nu)\,{\rm d}\nu = \sqrt{\nu\over 2\pi}\,
                              {\exp(-\nu/2)\over (2\pi)^{3/2} R_*^3}\,
                 {G(\gamma,\gamma\nu^{1/2})\over 2}\,{{\rm d}\nu\over\nu},
 \label{npkbbks}
\end{equation}
where the spectral parameters $R_*$ and $\gamma$ depend on the shape 
of the power spectrum of the initial density fluctuation field, whose 
definition is given in Appendix~B, along with that of the integral 
expression for the function $G(\gamma,\gamma\nu^{1/2})$. 
(Strictly speaking, BBKS considered density maxima rather than minima.  
However  Gaussian fluctuations are symmetric around the mean, so the 
density of peaks and troughs of the same absolute height is the same.)  

Notice that, in this model, the abundance of density minima in the 
primordial Universe depends on the depth of the minimum.   
If we define 
\begin{displaymath}
f(\nu)\,{\rm d}\nu \equiv (m/\bar\rho)\, n_{\rm v}(\nu)\,{\rm d}\nu
\end{displaymath}
and use the fact that the mass under a Gaussian filter is
\begin{equation}
m\,=\,{\bar\rho}\,(2\pi)^{3/2}R^3\,,
\end{equation}
then we have a quantity which one might interpret as the fraction of mass 
which is in minima of depth $\nu$. Unfortunately, for a comparison with 
the distribution of void sizes, this is a rather awkward quantity, 
since, in this picture, all voids contain the same mass $m$ whatever 
their height $\nu$ (because the smoothing radius $R$ is the same for 
all the voids).  

However, intuitively one would expect that deeper primordial minima 
should be identified with voids containing more mass, something which 
the above expression does not accomplish self-consistently. 
The model discussed in the next subsection attempts to account for the 
correlation between void mass and depth.  

\subsubsection{An Adaptive Troughs Model}
If, instead, we smooth the initial density field with a range of filter sizes 
$R$, and identify voids with minima of depth $\delta_{\rm v}/\sigma_0(m)$, 
then, because $\sigma_0$ decreases as $R\propto m^{1/3}$ increases, 
we have a model in which voids which contain more mass are associated 
with deeper minima. Appel \& Jones~\cite{appjon90} show how the 
changing smoothing scale modifies equation~(\ref{npkbbks}).  
The abundance of voids one obtains by replacing the BBKS~\cite{bbks} 
formula (our equation~\ref{npkbbks}) with the one given by 
Appel \& Jones~\cite{appjon90} is 
\begin{equation}
 \nu f(\nu) = \sqrt{\nu\over 2\pi}{\exp(-\nu/2)\over 3\,(R_*/R)^3}
            {H(\gamma,\gamma\nu^{1/2})\over \gamma\nu^{1/2}}
            {R^2\sigma_1^2(R)\over\sigma_0^2(R)}\,
            {{\rm d}m/m\over {\rm d}\nu/\nu},
 \label{fpkappeljones}
\end{equation}
in which we have set $m/\bar\rho = (2\pi)^{3/2}R^3$, the relation 
between mass and filter radius for a Gaussian smoothing filter, 
$\gamma$ is defined in equation~(\ref{bbksrg}), and 
\begin{displaymath}
 H(\gamma,y) = \int_0^\infty {\rm d}x\,x f(x)\,
             {\exp[-(x-y)^2/2(1-\gamma^2)]\over\sqrt{2\pi (1-\gamma^2)}},
\end{displaymath}
where $f(x)$ is given in equation~(\ref{bbksfx}), 
At large $\nu$ (i.e., for deep minima), $H\approx \gamma\nu^{1/2}\,G$, 
where $G$ is defined in Appendix~\ref{peaks}, so this expression is 
the same as equation~(\ref{npkbbks}).  
The two expressions differ significantly at smaller $\nu$.  
If the initial spectrum of density fluctuations was a power law, 
$P(k)\propto k^n$, then equation~(\ref{fpkappeljones}) for the 
void mass function becomes   
\begin{equation}
 \nu f(\nu) = \sqrt{\nu\over 2\pi}\exp\left(-{\nu\over 2}\right)
              {H(\gamma,\gamma\nu^{1/2})\over 2\ \gamma\nu^{1/2}}
              \left({5+n\over 6}\right)^{3/2},
 \label{fpkappeljones2}
\end{equation}
where we have used the fact that, for a Gaussian filter, 
$\gamma^2 = (n+3)/(n+5)$, and $(R/R_*)^2 = (n+5)/6$.  
Comparison with the excursion set approximation 
(equation~\ref{vfvapprox}) shows that both estimates contain 
the term $\sqrt{\nu/2\pi}\exp(-\nu/2)$, responsible for the  
exponential cut-off at large sizes. However, the additional 
correction factors differ substantially. For instance, in contrast 
to the excursion set formula, the correction factor in this primordial 
troughs model explicitly depends on the shape of the initial power 
spectrum.  

Although the distribution of void sizes associated with 
equation~(\ref{fpkappeljones}) cuts off exponentially at large sizes, 
as does the excursion set formula, it diverges at small sizes:  
\begin{equation}
n(m)\,\propto\,m^{-2} .
\end{equation}
Since this peaks model ignores both the void-in-void and the 
void-in-cloud processes, the divergence towards small void sizes is 
likely to be a significant overestimate. However, the large scale 
cut-off is likely to be accurate, probably even more-so than the 
excursion-set approximation (see below).  

\begin{figure}
\centering
\vskip -0.6truecm
\epsfxsize=9.8cm
\mbox{\hskip -0.6truecm\epsfbox{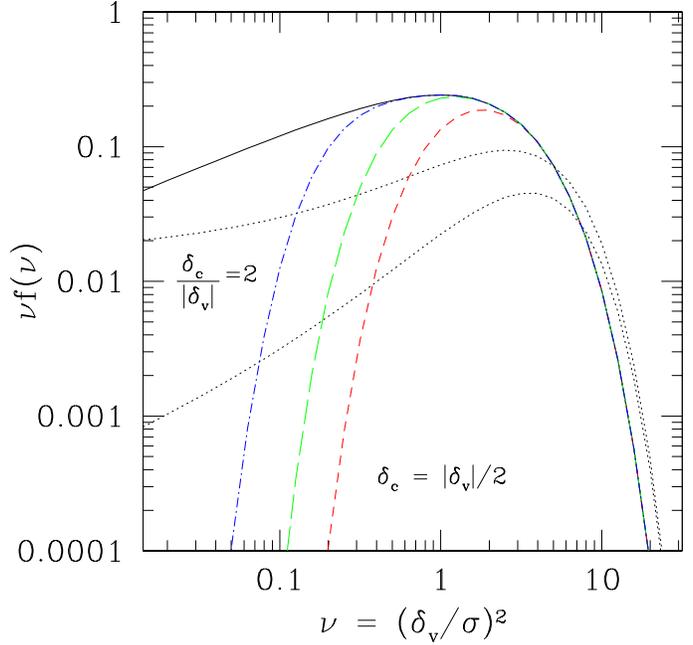}}
\vskip -0.5truecm
 \caption{Scaled distribution of void masses/sizes (equation~\ref{vfvoid}):  
  voids which contain large masses have large values of $\nu$.  
  Ignoring the {\em void-in-cloud} 
  process entirely yields the upper most curve. The spherical evolution 
  model suggests $\delta_{\rm c}\approx |\delta_{\rm v}|/2$; in this case, 
  the void distribution is reasonably well peaked about a characteristic 
  value.  Lower and upper dotted lines which extend to small values of 
  $\nu$ show predictions derived from equations~(\ref{npkbbks}) 
  and~(\ref{fpkappeljones}) of the peaks model.  
  These predictions depend on the shape of the initial power spectrum:  
  curves assume $P(k)\propto k^{-1.5}$.}
 \label{fvoid}
\end{figure}

For comparison, the lower and upper dotted curves in Fig.~\ref{fvoid} 
show the two predictions associated with these primordial troughs 
models:  equations~(\ref{npkbbks}) and~(\ref{fpkappeljones}). 
These predictions depend on the shape of the initial power spectrum, 
and for the curves in Fig.~\ref{fvoid} we have assumed 
$P(k)\propto k^{-1.5}$. The contrast between the small scale 
divergence of the peak/troughs formulae, with the small scale cut-off 
for the excursion set distributions, is obvious. 
Notice that the peaks/troughs models predict systematically more very 
large voids than does the excursion set model.  
The reason for this is closely related to the fact that the excursion 
set model does not include a factor like that in equation~(8) of 
Appel \& Jones~\cite{appjon90}.  For this reason, at large $\nu$, the 
peaks--based model is likely to be more accurate. 

\subsection{Void distribution and spatial patterns}
Our extension of the excursion set formalism provides a useful 
framework within which to construct an understanding of the 
dichotomy between the overdense and underdense regions of space in 
any hierarchical structure formation scenario. 

Because voids occupy most of the volume, the peaked void 
distribution predicted by our excursion analysis has strong 
implications for the expected spatial patterns in the cosmic matter 
distribution. Since the sizes of most voids will be similar to the 
characteristic void size, our findings suggest that the cosmic 
matter distribution will resemble a {\it foamlike packing of spherical 
voids of similar size and excess expansion rate}. 
The dynamical origin of such a matter distribution had been recognized 
by various authors, in particular within the context of analyses based 
upon an extrapolation of the Zel'dovich approximation (see Shandarin 
\& Zel'dovich~1989) and its extension, the adhesion 
approximation (Kofman, Pogosyan \& Shandarin~1990). 
The role of voids in the latter had indeed been recognized by 
Sahni, Sathyaprakash \& Shandarin~(1994). These studies 
spurred the concepts of a {\em cosmic web} or {\em cosmic skeleton} 
(see e.g. Van de Weygaert~1991, Bond, Kofman \& Pogosyan~1996, 
Novikov, Colombi \& Dor\'e~2003).  
Such patterns are naturally expected for cosmological scenarios with a 
lowpass power spectrum, characterized by a sharp spectral cutoff, as 
they would imply the imprint of an intrinsically dominating spatial 
scale.  
The {\em four mode excursion} formalism demonstrates and explains 
why the presence of such patterns is the natural outcome for a 
considerably wider range of Gaussian structure formation models. 

As an interesting thought experiment, suppose 
we extrapolate our findings to an ultimate and asymptotic extreme:  
What if we approximate the ``peaked'' void distribution by a 
``spiked'' distribution centered on the characteristic void size?  
In such a scenario, the cosmic matter distribution would be organized 
by a population of {\em equally sized}, spherical voids, all 
expanding at the same rate, akin to the scenario suggested by 
Icke~\cite{vi84}.  In this idealization, the walls and filaments would 
be found precisely at the midplanes between expanding voids, and the 
resulting skeleton of the matter distribution would be precisely that 
of a {\em Voronoi Tessellation} (Voronoi~1908; 
Okabe et al.~2000 and references therein). 
Our results appear to offer an explanation for the fact that heuristic 
models, based upon the use of tessellations as spatial templates for 
the galaxy distribution, can succesfully reproduce a variety of galaxy 
clustering properties (Van de Weygaert \& Icke~1989; Van de Weygaert~1991; 
Goldwirth et al.~1995).

\section{The Void Hierarchy}\label{voidhier}
The void distribution function derived in the previous section 
allows us to study in some detail the processes involved in the 
formation and development of void-dominated patterns in the 
cosmic matter distribution.  We have already discussed such 
gross features as the {\em void-filling factor}, and the {\em mass 
fraction} in voids.  But the excursion set analysis paves the way to 
a detailed assessment of the {\em temporal} dependence of a particular 
void, on its ``ancestral'' heritage as well as its {\em spatial} 
dependence on environmental factors. 
The following subsections touch upon a few of these elements 
of void evolution. 

\subsection{Void mass and volume fractions}
We have already argued that, in an Einstein de-Sitter universe, 
the mass fraction in voids does not evolve:  approximately one-third 
of the mass is in voids (equation~\ref{massfrac}), and that these 
voids fill space.  
This conclusion does not depend strongly on cosmological model.  
Because the collapse barrier $\delta_{\rm v}(a)$ decreases with time, 
the typical comoving void radius is larger at late times.  Therefore, 
the mass contained within a typical void is larger at late times.   
On the other hand, the total mass fraction does not evolve, 
from which we infer that the small mass voids present at early times 
must merge with each other to make the more massive voids which are 
present at later times.  

\subsection{Void ancestry}
The mass contained in a void at the present time was previously 
partitioned up among many smaller voids, each separated by their own 
walls. This distribution can be estimated similarly to how 
Bond et al.~\cite{bcek91} and Lacey \& Cole~\cite{lc93} estimate 
the growth of clusters. 

Consider a void $V_0$ which contains mass $M$ at a time $a_0$ 
when the critical densities for spherical collapse turnaround and 
void shell-crossing are $\delta_{\rm c0}$ and $\delta_{\rm v0}$, 
respectively. At an earlier epoch $a_1$, the critical densities 
were $\delta_{\rm c1}>\delta_{\rm c0}$ and 
$|\delta_{\rm v1}|>|\delta_{\rm v0}|$. 
The fraction of $M$ which was previously in voids that contained 
mass $m$ at the earlier time $a_1$ is given by inserting 
\begin{equation}
 \delta_{\rm c}\to\delta_{\rm c1}-\delta_{\rm v0};\quad 
 \delta_{\rm v}\to\delta_{\rm v1}-\delta_{\rm v0};\quad
  S\to S(m)-S(M)
\end{equation}
in equation~(\ref{vfvoid}). Integrating this over all possible 
ancestral voids (i.e., integrate over all $0< m \le M$), 
yields the mass fraction of $M$ which was in voids at the 
earlier epoch also:  
\begin{equation}
 f_{\rm void}(M)\,=\,1\,+\,\left(\delta_{\rm v1}-\delta_{\rm v0} \over 
                         \delta_{\rm c1}-\delta_{\rm v1}\right)\,
               =\,{\delta_{\rm c1}-\delta_{\rm v0} \over 
                        \delta_{\rm c1}-\delta_{\rm v1}}\,.
\end{equation}
Note how similar this expression is to the universal mass fraction 
in voids given by equation~(\ref{massfrac}).  
Note in particular that this fraction is less than unity. 
This reflects the fact that, at earlier times, some of the mass 
currently affiliated with the void $V_0$ was not part of 
the {\em ancestral} voids. Instead, this fraction of its matter 
content resided in the walls (and filaments) which partitioned 
$V_0$ into its many smaller constituent voids. 
In an Einstein--de-Sitter universe 
\begin{eqnarray}
\delta_{\rm c1}-\delta_{\rm c0} &=& \delta_{\rm c0}\,(z_1-z_0), 
\quad {\rm and}\nonumber\\
\delta_{\rm v1}-\delta_{\rm v0} &=& \delta_{\rm v0}\,(z_1-z_0)\,,
\end{eqnarray}
so that the mass fraction of void matter which was in voids at 
the earlier time also is 
\begin{equation}
f_{\rm void}(M)\,=\,1\,-\,{\cal D}_0\,{z_1-z_0 \over (1+z_1)}\,, 
\end{equation}
where 
${\cal D}_0 \equiv |\delta_{\rm v0}|/(\delta_{\rm c0}-\delta_{\rm v0})$ 
is the {\em void-and-cloud} parameter at the current epoch.   
Thus, at $z_1\approx z_0$, this fraction is close to unity, whereas 
for large lookback times $z_1 \gg z_0$ it tends to $1 - {\cal D}_0$
which is equal to the global void mass fraction 
(equation~\ref{massfrac}). In other words, the large voids emerging 
nowadays are to be traced back to an approximately average cosmic 
volume at early times. 

The transformations above allow one to write down the excursion set 
predictions for the rate at which smaller voids merge to make bigger 
ones. The calculation is analogous to the one used when estimating 
the merger rates of collapsed halos, and we will leave it for future 
work. In other words, one may reconstruct the ancestry of voids, 
the {\em void merger tree}, although this exercise will be complicated 
by the high rate of premature void mortality. 

\subsection{Environmental Dependence}
Suppose we evaluate the density field smoothed on a grid with cells 
of size $R$. The smoothed density will fluctuate from cell to cell. In the 
excursion set approach, we find that voids in denser cells 
1) are smaller, 2) have a narrower size distribution, and 3) account for a 
smaller fraction of total mass in the cell they inhabit.  
This subsection quantifies these trends of ``void bias''.   

Consider a cell of size $V$ within which the density is 
$\bar\rho (1+\delta)$; i.e., this cell contains mass 
$M=\bar\rho V(1+\delta)$. In the spherical evolution model, 
the initial and final densities are related:  
\begin{eqnarray}
 \delta_0(\delta)\!\!\! &=&\!\!\! {\delta_{\rm sc}\over 1.68647}
 \times \Biggl[1.68647 - {1.35\over (1+\delta)^{2/3}} \nonumber\\
                  &&\qquad\qquad\quad - {1.12431\over (1+\delta)^{1/2}} 
                  + {0.78785\over (1+\delta)^{0.58661}}\Biggr]
\label{d0mow}
\end{eqnarray}
(e.g. Mo \& White~1996). Note that $\delta_0$ has the same 
sign as $\delta$; initially dense regions become denser, whereas 
the comoving density in underdense regions decreases with time.  

In the context of the void model studied here, voids which are in 
cells of volume $V$ within which the overdensity is $\delta$ are 
described by random walks which do not start from the origin 
$[S=0,\delta_0=0]$, but from the position $[S(M),\delta_0(\delta)]$. 
Therefore the fraction of the total mass $M=\bar\rho V(1+\delta)$ 
which is in voids of mass $m$ is given by setting 
\begin{eqnarray}
 \delta_{\rm c}&\to&\delta_{\rm c}-\delta_0({\delta}),\qquad 
 \delta_{\rm v} \to \delta_{\rm v}-\delta_0({\delta})\nonumber\\
             S & \to &S(m)-S\Bigl[\bar\rho V(1+\delta)\Bigr]
 \label{deltatrans}
\end{eqnarray}
in equation~(\ref{vfvoid}). Integrating the resulting distribution over 
$0\le m\le M$ yields the fraction of mass, in a region of volume $V$ 
within which the density is $\delta$, which is contained in voids:  
\begin{equation}
 f_{\rm void}(\delta)\,=\,{\delta_{\rm c}-\delta_0(\delta) \over 
                           \delta_{\rm c}-\delta_{\rm v}}\,.  
\end{equation}
This indicates that the mass fraction $f_{\rm void}(\delta)$ 
decreases as the density $\delta$ of the cell increases.  Conversely, 
as $\delta\to -0.8$, the density we associate with a void, then 
$\delta_0(\delta)\to\delta_{\rm v}$, and so $f_{\rm void}(\delta)\to 1$ 
as expected.  (In this extreme, the fitting formula~(\ref{d0mow}) is 
slightly inaccurate, since it sets $\delta_0(-0.8)=-2.7$, rather than 
$-2.81$.)  
Thus, our analysis allows one to quantify a fact which is intuitively 
obvious: that dense regions have a smaller fraction of their mass 
in voids.  

Furthermore, the typical void size scales as  
\begin{equation}
S(m)\,\approx\,S(M)\,+\,|\delta_{\rm v}-\delta_0(\delta)|\,,
\end{equation}
where the void size $R(m)$ decreases as $S(m)$ increases.  
Since $|\delta_{\rm v}-\delta_0(\delta)|$ increases as $\delta$ 
increases, the typical void size is larger in regions of lower 
density.  Moreover, the sharpness of the peak in the void size 
distribution depends on $\delta_{\rm c}/|\delta_{\rm v}|$ 
(c.f. Fig.~\ref{fvoid}:  the void size distribution becomes more 
more sharply peaked as {\em void-in-cloud} demolition becomes 
more important.  The transformations in equation~(\ref{deltatrans}) 
mean that, in dense regions ($\delta>0$), where voids are more 
likely to be demolished by collapsing clouds, the distribution of 
void sizes is expected to be narrower.

\subsection{Spatial Clustering}
The model developed here also allows us to build an approximate 
model of the evolution of the dark matter correlation function 
following methods outlined in Neyman \& Scott~\cite{ns1952} 
and Scherrer \& Bertschinger~\cite{scherredb91} (recently 
reviewed by Cooray \& Sheth~2002).  
The calculation requires estimates of 
 1) the distribution void sizes, 
 2) the clustering of void centres on large scales, and 
 3) the density run within a void.  
The previous sections derived estimates for the first of these three 
quantities.  

The second one, the clustering of void centres, can be estimated as 
follows.  Write the two-point correlation function of voids which 
contain mass $m_1$ and $m_2$ as 
\begin{equation}
 \xi_{\rm vv}(r|m_1,m_2) = b(m_1)b(m_2)\,\xi_{\rm dm}(r),
\end{equation}
where $\xi_{\rm dm}$ is the correlation function of the dark matter, 
and the bias factor $b(m)$ depends on the mass or size of the voids.  
Following Cole \& Kaiser~(1989), Mo \& White~\cite{mw96} and 
Sheth \& Tormen~\cite{st99}, knowledge of the number density of 
objects is sufficient for estimating their spatial distribution, 
at least on large scales.  
Therefore, $b(m)$ depends on which estimate of $n_{\rm v}(m)$ we use.  
If we use equation~(\ref{fpkappeljones}) from the peaks model, then 
\begin{equation}
 b(m) = 1 + {\nu + h_1\over |\delta_{\rm v}|},
\end{equation}
where 
\begin{equation}
 h_k = {(-1)^k\over k!}
 {(\gamma \nu^{1/2})^k\over H(\gamma,\gamma\nu^{1/2})}
 {\partial^k H(\gamma,x)\over\partial x^k}
 \Biggr|_{x=\gamma\nu^{1/2}}.
\end{equation}
Using our approximation to the excursion set prediction 
(equation~\ref{vfvapprox}) instead gives 
\begin{equation}
 b(m) \approx 1 + {\nu-1\over |\delta_{\rm v}|} - 
                  {(\delta_{\rm v}/\delta_{\rm c})^2\over 
                    4\nu\, (\delta_{\rm c}+|\delta_{\rm v}|)}.
\end{equation}
In both cases, the largest voids are more strongly clustered than 
those of average size.  The higher order moments of the void 
distribution can be estimated similarly to how 
Mo, Jing \& White~\cite{mjw97} estimate the higher order moments 
of clusters.  

If we suppose that all the mass is contained in the void walls, 
then we can approximate the density run around a void centre as 
a uniform density shell.  
Figures~3 and~5 in Dubinski et al.~\cite{ddglp93} 
suggest this is a fair approximation.  Specifying the mass associated 
with the void as well as the shell thickness sets the density within 
the shell. Thus, we have all three ingredients required to model the 
power spectrum (or correlation function) of the dark matter 
distribution.  

There is one important aspect in which this void-based model for the 
correlation function differs from the usual halo-based model.  Namely, 
in the halo model, halos are treated as hard spheres which do not 
overlap; this leads to exclusion effects on small scales.  Since the 
radius of a typical collapsed halo is smaller than a Megaparsec, the 
effects of exclusion are expected to be unimportant.  
In a void-based model, on the other hand, typical void radii are of 
order a few Megaparsecs; since voids do not overlap, exclusion effects 
are likely to matter on scales of order a few Megaparsecs.  
We leave a more extensive analysis of all this to future work.

\section{Summary and Interpretation}\label{discuss}
Initially underdense regions expand faster than the Hubble flow.  
If they are not embedded within overdense regions, such regions 
eventually form voids which are surrounded by dense void walls. 
These voids expand with respect to the background Universe, and 
during their expansion tend to become more and more spherical 
(Figure~\ref{voidcdm6slc}).  The outward expansion is differential, 
so most initial void configurations tend to evolve to distinct 
``tophat'' density profiles (Figure~\ref{sphmodel}). 
A description of the evolution of initially spherical tophat over- 
and underdense regions has been available for some time 
(Appendix~\ref{stophat}). 
Although the spherical evolution model allows one to study the 
evolution of single isolated objects, a more complete theory must 
also describe void evolution within the context of a generic random 
density fluctuation field. 

The evolving void hierarchy is determined by two processes:  
\begin{itemize}
 \item The {\em void-in-void} process describes the evolution of a 
     system of voids which are embedded in a larger scale underdensity; 
     in this case small voids from an early epoch merge with one another 
     to form a larger void at a later epoch (Fig.~\ref{voidcdm6slc}). 
 \item The {\em void-in-cloud} process is associated with underdense 
     regions embedded within a larger overdense region; in this case 
     the smaller voids from an earlier epoch may be squeezed out of 
     existence as the overdense region around them collapses 
     (Fig.~\ref{voidcollnbody}). 
\end{itemize}
In contrast, the evolution of overdensities is governed only by the 
{\em cloud-in-cloud} process; the {\em cloud-in-void} process is much 
less important, because clouds which condense in a large scale void are 
not torn apart as their parent void expands around them.  

This asymmetry between how the surrounding environment affects halo and 
void formation can be incorporated into the {\em excursion set approach} 
by using one barrier to model halo formation and a second barrier to 
model void formation (Fig.~\ref{excurcensus}).  
Only the first barrier matters for halo formation, but both 
barriers play a role in determining the expected abundance of voids.  
The resulting void size distribution is a function of two parameters 
(equation~\ref{vfvoid}), which the model associates with the dynamics 
of expansion and collapse. 
The predicted distribution of voids is well-peaked about a 
characteristic size (Figs.~\ref{fvlin} and~\ref{frvoid})---in contrast, 
the distribution of halo masses is not.  
Comparison of the two-parameter family of void distribution curves 
(Figure~\ref{fvoid}) with the void size distribution in numerical 
simulations of hierarchical clustering is the subject of work in 
progress (Colberg et al. 2004). 

Five major observations about the properties of the void population 
result from the {\em two-barrier} excursion set model: 
\begin{itemize}
\item{} The {\em void-in-cloud} mechanism (Fig.~\ref{voidcollnbody}) 
  is responsible for the demise of a sizeable population of small 
  voids.  As a result, the void size distribution has a small-scale 
  cut-off:  the void population is ``void'' of small voids 
  (Section~\ref{voiddistr}), in a way which our excursion set 
  analysis quantifies.  
\item{} The population of large voids is insensitive to this effect
  (Fig.~\ref{fvlin}).  Therefore, the abundance of voids which are 
  larger than the typical initial comoving sizes of clusters should 
  be well described by peaks theory or its extensions described in 
  Section~\ref{ajpeaks}.  
\item{} At any cosmic epoch there is a {\em characteristic void size} 
  which increases with time: the larger voids present at late times 
  formed from mergers of smaller voids which formed at earlier times 
  (e.g Fig.~\ref{voidcdm6slc} and Section~\ref{voidhier}).  
\item{} At any given time the mass fraction in voids is approximately 
  thirty percent of the mass in the Universe, and the voids 
  approximately fill space (Section~\ref{voidhier}). 
\item{} As the size of most voids will be similar to the
  characteristic void size, the cosmic matter distribution resembles 
  a {\em foamlike packing of spherical voids of approximately similar 
  size and excess expansion rate}. This may explain why simple models 
  based on the Voronoi tesselation exhibit many of the features so 
  readily visible in N-body simulations of hierarchical clustering.  
\end{itemize}

\subsection{Galaxies in Voids}
It is with some justification that most observational attention is 
directed to regions where most of the matter in the Universe has 
accumulated. Almost by definition they are the sites of most 
observational studies, and the ones that are most outstanding in 
appearance. Yet, for an understanding of the formation of the large 
coherent foamlike patterns pervading the Universe, it may be well 
worth directing attention to the complementary evolution of underdense 
regions. These are the progenitors of the observed voids, the vast 
regions in the large-scale cosmic galaxy distribution that are 
practically devoid of luminous matter. 

When extensive systematic redshift surveys began mapping the spatial 
galaxy distribution, voids were amongst the most visually striking 
features.  Since then, the role of voids as key ingredients of the 
cosmic galaxy distribution has been demonstrated repeatedly in 
extensive galaxy redshift surveys (see Kauffmann \& Fairall~1991; 
El-Ad, Piran \& da Costa~1996; El-Ad \& Piran~1997; 
Hoyle \& Vogeley~2002; Plionis \& Basilakos~2002; Rojas et al. 2003). 
A number of studies also indicated that observed voids exhibit 
distinct hierarchical features. Van de Weygaert~(1991) suggested 
the existence of a {\it ``void hierarchy''} when pointing out that 
the galaxy distribution in the CfA/SRSS2 redshift survey 
(Geller \& Huchra~1989; da Costa et al.~1993) gave the impression of 
small-scale voids embedded in the less pronounced large-scale 
underdense region delimited by the ``Great Wall''. 
Even in the most canonical specimen amongst its peers, the Bo\"otes 
void, traces of a faint structured internal galaxy distribution were 
found (Szomoru et al.~1996).  

The dynamical impact of voids has proven to be crucial for 
understanding the cosmic flow patterns in the Local Universe. 
Measured peculiar galaxy velocities imply reconstructions of the 
local cosmic density field in which the repulsive actions of voids 
are important (e.g. Bertschinger et al.~1990; 
Strauss \& Willick~1995; Dekel \& Rees~1995). 
And more locally, the void's influence on cosmic flows was 
established when Bothun et al.~\cite{bothun92} studied galactic 
peculiar motions along a wall around the largest void in the CfA 
redshift sample (de Lapparent, Geller \& Huchra~1986). 

In all these respects, voids in the galaxy distribution are similar 
to those in the dark matter distribution.  However, 
although voids in the galaxy distribution are mostly roundish in 
shape, they have typical sizes in the range of 
$20h^{-1}-50h^{-1}$Mpc (e.g. Hoyle \& Vogeley~2002; 
Plionis \& Basilakos~2002; Arbabi-Bidgoli \& M\"uller~2002). 
These sizes are considerably in excess of the typical void diameters 
in our model of voids in the dark matter distribution, but note that 
the typical void size in the galaxy distribution depends on the galaxies 
which were used to define the void.  The voids associated with rare 
luminous galaxies are larger in part because the number density of 
such galaxies is lower.  As we describe below, our excursion set 
analysis provides a framework for modeling this dependence.  

In recent years, the possibility that void galaxies are a 
systematically different population has received considerable 
attention (see e.g. Szomoru et al.~1996; El-Ad \& Piran~2000; 
Peebles~2001; Mathis \& White~2002, Rojas et al.~2003; 
Benson et al.~2003). 
In the simplest models of biased galaxy formation 
(e.g. Little \& Weinberg~1994) one would expect to find 
voids filled with galaxies of low luminosity, or galaxies of some 
other uncommon nature (e.g. Hoffman, Silk \& Wyse~1992). 
Indeed, even though various studies were oriented towards establishing 
the properties of voids in galaxies (e.g., Kauffmann \& Fairall~1991; 
El-Ad \& Piran~1997,~2000; Hoyle \& Vogeley~2002; 
Arbabi-Bidgoli \& M\"uller~2002; Plionis \& Basilakos~2002), 
and some focussed explicitly on the identity of galaxies inside voids 
(e.g. Szomoru et al.~1996; El-Ad \& Piran~2000; Rojas et al.~2003), 
a clear picture of the relation between void galaxies and their 
surroundings is only just becoming available.  
This is in large part due to the fact the large scale surveys such 
as the SDSS (Abazajian et al.~2002) and 2dFGRS (Colless et al.~2003)
now probe a sufficiently large cosmological volume that they contain 
a statistically significant number of large voids.  

Recently, Mathis \& White~\cite{mathwhit2002} and 
Benson et al.~\cite{bensetal2003} have identified and studied voids 
and void galaxies in semi-analytic galaxy formation models. 
In these models, the properties of galaxies are determined by the halos 
they inhabit.  Therefore, if one can model the halo population associated 
with voids, a model of the void galaxy population is within reach.  
The excursion set model developed here is phrased in the same language 
used in the simulations, so it represents the ideal framework within 
which to attempt such a model.  

In particular, consider the {\em cloud-in-void} process shown in the 
second series of panels in Fig.~\ref{excurcensus}.  
Notice that the condition that the cloud exist in a void means that, 
on average, clouds in voids will be less massive than clouds in 
regions of average density (to represent a cloud, the walk must 
reach $\delta_{\rm c}$, and on average, it will take more steps to 
travel to $\delta_{\rm c}$ from $\delta_{\rm v}$ than from 
zero---more steps imply smaller masses).  For similar reasons, 
the clouds associated with the more massive halos should be more 
massive on average (this is also discussed more fully by 
Mo \& White~1996; Sheth \& Tormen~2002 and Gottl\"ober et al.~2003).  
Although we speak of the clouds as being within the voids, our 
discussion of how voids empty their mass into the ridge which 
surrounds them (c.f. Fig.~\ref{sphmodel}) suggests it may be more 
appropriate to think of these clouds as being associated with the 
void walls.  
It seems natural to associate void galaxies with such clouds-in-voids.  
If low mass halos host lower mass galaxies, and less massive galaxies 
tend to be less luminous and bluer, then void galaxies should be 
fainter and bluer than field or cluster galaxies; our model allows 
one to quantify this trend.  Thus, the results presented here allow 
a more elaborate model for voids in the galaxy distribution
and the galaxy population in voids than that discussed 
recently by Friedmann \& Piran~\cite{frpir2001}.  
Developing such a model is the subject of work in progress.

\section*{acknowledgements}
We thank J\"org Colberg, Bernard Jones and Paul Schechter for 
encouraging discussions and suggestions, and A. Babul for useful 
conversations on collapsing voids. 
This work was supported in part by the DOE and NASA grant 
NAG 5-10842 at Fermilab.  

\noindent O, what men dare do! What men may do! 
What men daily do, not knowing what they do! 
(Shakespeare 1598)

\appendix
\section{The Spherical Tophat Model}\label{stophat}
\subsection{Background}
Analytically tractable idealizations help to understand various 
aspects of void evolution. In this regard, the {\em spherical model} 
represents the key reference model against which we may assess the 
evolution of more complex configurations. Also, it provides the 
clearest explanation for the various void characteristics listed 
in the main test. And most significantly within the context of this 
work, it provides the fundament from which our formalism for 
hierarchical void evolution is developed. 

The structure of a spherical void or peak can be treated in terms of 
mass shells. In the ``spherical model'' concentric shells remain 
concentric and are assumed to be perfectly uniform, without any 
substructure. The shells are supposed never to cross until the 
final singularity, a condition whose validity is determined by the 
initial density profile. The resulting solution of the equation 
of motion for each shell may cover the full nonlinear evolution 
of the perturbation, as long as shell crossing does not occur. 

The treatment of the spherical model in a cosmological context 
has been fully worked out (Gunn \& Gott~1973; 
Lilje \& Lahav~1991).  
As long as the mass shells do not cross, they behave as 
mini-Friedmann universes whose equation of motion assumes exactly the 
same form as that of an equivalent FRW universe with a modified value 
of $\Omega_s$. The details of the distribution of the mass interior to 
the shell are of no direct relevance to the evolution of each 
individual shell. Instead, the evolution depends on the total 
mass contained within the radius of the shell. and the global 
cosmological background density.

Although quantitative details depend on the cosmological model, 
a study of the evolution of spherical perturbations in an 
Einstein-de Sitter Universe suffices to illustrate all the 
important physical features.  

\subsection{Definitions}
When a mass shell at some initial time $t_i$ starts expanding from 
a physical radius $r_i=a(t_i) x_i$, its subsequent motion is 
characterized by the expansion factor ${\cal R}(t,r_i)$ of the shell: 
\begin{equation}
 r(t,r_i)\,=\,{\cal R}(t,r_i) r_i\,,
\end{equation}
\noindent where $r(t,r_i)= a(t) x(t,x_i)$ is the physical radius of the shell 
at time $t$ and $x(t,x_i)$ the corresponding comoving radius. The evolution 
of the shell is dictated by the cosmological density parameter 
\begin{equation}
 \Omega(t)\,={\displaystyle \,8\pi G\,\rho_u(t)\over \displaystyle 3 H_u^2}\, 
\end{equation}
and the mean density contrast within the radius of the shell, 
\begin{eqnarray}
 \Delta(r,t)&\,=\,&{\displaystyle 3\over \displaystyle r^3 }\int^r_0
 \left[{\displaystyle \rho(y,t)\over\displaystyle{\rho_u}(t)}-1\right]\,
  y^2\,dy \,\nonumber\\
 &\,=\,&{\displaystyle 3 \over \displaystyle r^3} \int^r_0 \delta(y,t)\, y^2\,dy\,,
\end{eqnarray}
To determine the evolution of $R(t,r_i)$, it is convenient to introduce 
the parameters $\Delta_{ci}=\Delta_c(t_i)$ and $\alpha_i$ where
\begin{eqnarray}
 1+\Delta_{ci} &\,=\,& \Omega_i \left[1+\Delta(t_i,r_i)\right]\nonumber\\
      \alpha_i &\,=\,& \left({\displaystyle v_i \over \displaystyle H_i r_i}\right)^2 -1\,.
\end{eqnarray}
Here, $\Omega_i=\Omega(t_i)$, $H_i=H(t_i)$ and 
$v_i$ is the physical velocity (i.e. the sum of the peculiar 
velocity and Hubble expansion velocity with respect to the void 
center) of the mass shell at $t=t_i$. 
The usual assumption of a growing mode perturbation implies that 
the velocity perturbation $v_{pec,i}$ for a spherical perturbation, 
at the initial time $t_i$, is 
\begin{equation}
 v_{pec,i}\,=\,-{\displaystyle H_ir_i \over \displaystyle 3}
 f(\Omega_i) \Delta(r_i,t_i)\,,
\end{equation}
\noindent and hence,
\begin{equation}
 \alpha_i\,=\,-{2\over 3} f(\Omega_i) \Delta(r_i,t_i)\,.
\end{equation}
In effect, $\Delta_{ci}$ is the density contrast of the shell 
with respect to a critical universe ($\Omega=1$) at the cosmic time 
$t_i$, while $\alpha_i$ is a measure of the corresponding peculiar 
velocity (or, rather, the kinetic energy) of the shell. 
The evolution of a spherical over- or underdensity is {\it entirely} 
and {\it solely} determined by the initial (effective) over- or 
underdensity within the (initial) radius $r_i$ of the shell, 
$\Delta_{ci}(r_i,t_i)$, and the corresponding velocity perturbation, 
$v_{pec,i}$. Hence, the values of $\Delta_{ci}$ and $\alpha_i$ 
determine whether a shell will stop expanding or not, i.e. whether 
it is closed, critical or open. The criterion for a closed shell is 
$\Delta_{ci}>\alpha_i$, for a critical shell, $\Delta_{ci}=\alpha_i$, 
and $\Delta_{ci}<\alpha_i$ for an open shell. 

Notice that these expressions assume that the initial density 
fluctuation was negligible, so that the initial mass $m$ and 
initial comoving size $R$ are related: $m\propto R^3$.  

\subsection{Shell Solutions}\label{sphsol}
The solution for the expansion factor 
${\cal R}(t,r_i)={\cal R}(\Theta_r)$ of an overdense cq. underdense 
shell is given by the parametized expressions 
\begin{equation}
 {\cal R}(\Theta_r)\,=\,\cases{{1 \over 2}{\displaystyle 1 + \Delta_{ci} \over \displaystyle   
(\alpha_i-\Delta_{ci})}\,\,\left(\cosh \Theta_r - 1\right) & \quad $\Delta_{ci}<\alpha_i$,\cr
\cr\cr
{1 \over 2}\,{\displaystyle 1 + \Delta_{ci} \over \displaystyle (\Delta_{ci}-\alpha_i)}\,\,
\left(1-\cos \Theta_r\right)&\quad $\Delta_{ci}>\alpha_i$\,,\cr}
\end{equation}
\noindent in which the development angle $\Theta_r$, 
which paramaterizes all physical quantities relating to the mass shell, 
is related to time $t$ via
\begin{equation}
 t(\Theta_r)\,=\,\cases{{1 \over 2}\,{\displaystyle 1 + \Delta_{ci} \over \displaystyle   
(\alpha_i-\Delta_{ci})^{3/2}}\,\,\left(\sinh \Theta_r - \Theta_r\right) & \quad $\Delta_{ci}<\alpha_i$,\cr
\cr\cr
{1 \over 2}\,{\displaystyle 1 + \Delta_{ci} \over \displaystyle (\Delta_{ci}-\alpha_i)^{3/2}}\,\,
\left(\Theta_r-\sin \Theta_r\right)&\quad $\Delta_{ci}>\alpha_i$\,,\cr}
\end{equation}
\noindent while for a critical shell the solution is given by the 
direct relation
\begin{equation}
 {\cal R}(\Theta_r)\,=\,\left\{{3 \over 2} H_i (1+\Delta_{ci})^{1/2} t\right\}^{2/3} \quad\quad\quad \Delta_{ci}=\alpha_i\,.
\end{equation}
\noindent Notice that the solutions for the evolution of overdense 
and underdense regions in essence are the same, and are interchangeable 
by replacing 
\begin{eqnarray}
(\sinh \Theta - \Theta) &\quad\Rightarrow\quad& (\Theta - \sin\Theta)\nonumber\\ 
(\cosh \Theta-1) &\quad\Rightarrow\quad& (1-\cos \Theta)\,.
\end{eqnarray}

\subsection{Density Evolution}\label{sphden}
If the initial density contrast of a shell is $\Delta_i(r_i)$, 
its density contrast $\Delta(r,t)$ at any subsequent time $t$ is given by 
\begin{equation}
1+\Delta(r,t)\,=\,{\displaystyle 1+\Delta_i(r_i) \over \displaystyle {\cal R}^3}\,
{\displaystyle a(t)^3 \over \displaystyle a_i^3}\,,
\end{equation}
\noindent With $\Delta(r,t)$ being a relative quantity, comparing 
the density of the mass shell at radius $r$ at time $t$ with that of 
the global cosmic background, the value of $\Delta(r,t)$ is a function 
of the shell's development angle $\Theta_r$ as well as that of the 
development angle of the Universe $\Theta_u$, 
\begin{equation}
\Omega=\cases{{\displaystyle 2 \over \displaystyle \cosh \Theta_u + 1} 
&\quad\quad\quad $\Omega < 1$,\cr
\cr
{\displaystyle 2 \over \displaystyle \cos \Theta_u + 1}&\quad\quad\quad $\Omega > 1$.\cr}
\end{equation}
The shell's density contrast may then be obtained from  
\begin{equation}
 1+\Delta(r,t) = f(\Theta_r)/f(\Theta_u),
\end{equation}
where $f(\Theta)$ is the cosmic ``density'' function:
\begin{equation}
 f(\Theta)\,=\,\cases{{\displaystyle \left( \sinh \Theta - \Theta \right)^2 \over 
\displaystyle  \left( \cosh \Theta - 1 \right)^3} & \qquad {\rm open} \,,\cr
\cr
{2/9} & \qquad {\rm critical}\,,\cr
\cr
{\displaystyle \left( \Theta - \sin \Theta \right)^2 \over 
\displaystyle \left( 1 - \cos \Theta \right)^3}& \qquad {\rm closed}\,,\cr}.
\end{equation}
This expression is equally valid for the shell 
(in which case ``open'' means $\Delta_{ci}<\alpha_i$) 
and the global background Universe (where ``open'' means $\Omega<1$). 

\subsection{Shell Velocities}\label{sphvel}
The velocity of expansion or contraction of a spherical shell 
is given by computing ${\rm d}{\cal R}/{\rm d}t$, so it can be 
written in terms of $\Theta_r$ and $\Theta_u$.  In particular, 
the shell's peculiar velocity with respect to the global Hubble 
velocity, 
\begin{equation}
 v_{pec}(r,t)\,=\,v(r,t)-H_u(t) r(t)\,,
\end{equation}
may be inferred from the expression
\begin{equation}
 v_{pec}(r,t)\,=\,H_u(t) r(t)\,\left\{{g(\Theta_r) \over g(\Theta_u)}
\,-\,1\right\}\,,
\end{equation}
where $H_u(t)r =({\dot a}/a)r$ and the cosmic ``velocity'' function is 
\begin{equation}
 g(\Theta)\,=\,\cases{{\displaystyle \sinh \Theta \left( \sinh \Theta - \Theta \right)
\over \displaystyle  \left( \cosh \Theta - 1 \right)^2} & \qquad {\rm open}\,,\cr
\cr
{\displaystyle 2 \over \displaystyle 3} & \qquad {\rm critical}\,,\cr
\cr
{\displaystyle \sin \Theta \left( \Theta - \sin \Theta \right) \over 
\displaystyle \left( 1 - \cos \Theta \right)^2}& \qquad {\rm closed}\,.\cr}
\end{equation}
Thus, we may define a Hubble parameter $H_s$ for each individual shell, 
\begin{equation}
 H_s(r,t)\,\equiv\,{\displaystyle {\dot{\cal R}}\over\displaystyle {\cal R}}\,=
\,H_u(t)\,\left\{{\displaystyle g(\Theta_r)\over\displaystyle g(\Theta_u)}\right\}\,.
\end{equation}

\subsection{Overdensities and collapse when $\Omega=1$}\label{sphcol}
The previous sections provided explicit expressions for the evolution 
of a spherical perturbation in FRW backgrounds with no cosmological 
constant.  To better illustrate our argument, we will now specialize 
to the case of an Einstein de-Sitter model.  It will prove useful 
to contrast the spherical evolution with that predicted by linear 
theory.  We will use $D(z)$ to denote the {\it linear density 
perturbation growth factor}, normalized so that $D(z=0)=1$.
For an Einstein-de Sitter Universe,$D(z)=1/(1+z)$.  
Note that this makes $D\propto (t/t_0)^{2/3}$.  
Similarly, the growth of velocities in linear theory is given by 
\begin{equation} 
v_{\rm lin}(r) \,=\, -{H_u r\over 3}\,f(\Omega)\,\Delta(r,t)\,,
\end{equation}
where $f(\Omega)\approx \Omega^{0.6}$ (Peebles~1980). 
It is a useful exercise to verify that, in its early stages (i.e., 
small development angle), the spherical evolution model does indeed 
reproduce linear theory.  

Consider the evolution of an initially overdense (or, rather, bound) shell.  
Such a shell will initially expand slightly slower than the background, 
this expansion gradually slowing to a complete halt, after which it 
turns around and starts to contract. 
At turnaround, $v(r,t)=0$, so $\Theta_r=\pi$, and the density is 
\begin{equation}
 1 + \Delta(r,t_{ta}) = (3\pi/4)^2.
\end{equation}
Therefore, at turnaround, the {\it comoving radius} of a spherical 
perturbation has shrunk by a factor of $(3\pi/4)^{2/3}=1.771$ from 
what it was initially.  
Had the perturbation evolved according to linear theory, then 
turnaround would happen at that redshift when the linear theory 
prediction $\Delta_{lin}$, reaches the value $\delta_{\rm ta}$:
\begin{equation}
\Delta_{lin}(z_{\rm ta}) = \delta_{\rm ta}
                         = (3/5)(3\pi/4)^{2/3}\approx 1.062.
\end{equation}
Full collapse is associated with $\Theta_r=2\pi$.  
At this time, the linearly extrapolated initial overdensity reaches 
the threshold value $\delta_{\rm c}$, 
\begin{equation}
\Delta_{lin}(z_{\rm c}) = \delta_{\rm c}  
               = \left(3 \over 5\right) \left(3\pi \over 2\right)^{2/3}\,
               \approx\,1.686\,. 
\end{equation}
This makes it straightforward to determine the collapse redshift 
$z_{\rm coll}$ of each bound perturbation directly from a given 
initial density field. In terms of the primordial field linearly 
extrapolated to the present time, $\Delta_{lin,0}$, the collapse 
redshift $z_{\rm coll}$ may be directly inferred from 
\begin{equation}
 D(z_{\rm coll})\,\Delta_{lin,0}\,=\,\delta_{\rm c}\ . 
\end{equation}
so  
\begin{equation}
1+z_{coll}\,=\,{\displaystyle \Delta_{lin,0}\over 1.686}.
\end{equation}
Formally, at collapse, the comoving radius is vanishingly small 
(${\cal R}(2\pi)=0$).  In reality, the matter in the collapsing object 
will virialize as interactions between matter in the shells will 
exchange energy betwen the shells and ultimately an equilibrium 
distribution will be found.  Therefore, it is usual to assume that 
the final size of a collapsed spherical object is finite and equal 
to its virial radius. For a perfect {\it tophat} density, the 
object's final size $R_{\rm fin}$ is then $\approx 5.622$ times 
smaller than it was initially (Gunn \& Gott~1973), i.e. 
\begin{equation}
 R_{\rm fin}/{\widetilde R}_{i,coll}\,=\,(18\pi^2)^{1/3}\,\approx 5.622\,,
\end{equation}
where ${\widetilde R}_{i,coll}\equiv R_i (a_{coll}/a_i)$.  

\subsection{Underdensities and shell-crossing when $\Omega=1$}\label{sphexp}
Underdense spherical regions evolve differently than their 
overdense peers. The outward directed peculiar acceleration is 
directly proportional to the integrated density deficit $\Delta(r,t)$ 
of the void. In the generic case, the inner shells ``feel'' a stronger 
deficit, and thus a stronger outward acceleration, than the outer shells. 

Once again, to better illustrate our argument, we will now specialize to 
the case of an Einstein de-Sitter model.  The density deficit evolves as 
\begin{equation}
 1+\Delta(r,z) \approx {9 \over 2}\,{(\sinh \Theta_r - \Theta_r)^2 \over (\cosh \Theta_r-1)^3}.
\end{equation}
In comparison, the corresponding linear initial density deficit 
$\Delta_{lin}(z)$:
\begin{equation}
\Delta_{lin}(z) = {\displaystyle \Delta_{lin,0} \over \displaystyle 1+z}
\approx -{3 \over 4}^{2/3}\,{\displaystyle \left(\sinh \Theta_r - \Theta_r\right)^{2/3} 
\over \displaystyle {5/3}}\,.
\end{equation}
The (peculiar) velocity with which the void expands into its surroundings is 
\begin{equation}
v_{pec}(r,t) \,=\,H_u r\, \left\{{3\over 2}
{\displaystyle \sinh\Theta_r\,(\sinh \Theta_r-\Theta_r) \over \displaystyle (\cosh\Theta_r-1)^2}\,-\,1\right\}\,.
\end{equation}

As a consequence of the differential outward expansion within 
and around the void, and the accompanying decrease of the expansion 
rate with radius $r$, shells start to accumulate near the boundary 
of the void. The density deficit $|\Delta(r)|$ of the void decreases 
as a function of radius $r$, down to a minimum at the center. 
Shells which were initially close to the centre will ultimately catch 
up with the shells further outside, until they eventually pass them. 
This marks the event of {\it shell crossing}. 
The corresponding gradual increase of density will then have turned 
into an infinitely dense ridge.  
From this moment onward the evolution of the void may be described 
in terms of a self-similar outward moving shell 
(Suto et al.~1984; Fillmore \& Goldreich~1984; Bertschinger~1985). 
Strictly speaking, this only occurs for voids whose density profile 
is sufficiently steep, since a sufficiently strong differential 
shell acceleration must be generated.  
This condition is satisfied at the step-function density profile 
near the edge of a tophat void. 

For a tophat void in an Einstein-de Sitter Universe the shells initially 
just outside the void's edge pass through a shell crossing stage at a 
precisely determined value of the mass shell's development angle 
$\Theta_r=\Theta_{sc}$, 
\begin{equation}
 {\sinh \Theta_{sc} \left( \sinh \Theta_{sc} - \Theta_{sc} \right)
   \over \left(\cosh \Theta_{sc} - 1 \right)^2} = {8\over 9},
   \quad {\rm so}\ \Theta_{sc} \approx 3.53.
\end{equation}
At this shell-crossing stage, the average density within the void is 
\begin{equation}
 1+\Delta(r,t)\,=\,0.1982 
\end{equation}
times that the cosmic background density. 
This means that the shell has expanded by a factor of 
$(0.1982)^{-1/3}\approx 1.7151$.  
In comparison, the underdensity estimated using linear theory 
at the time of shell crossing is 
\begin{equation}
\Delta_{lin}(z_{\rm sc}) = \delta_{\rm v} = -\left(3 \over 4\right)^{2/3}\,
      {\left(\sinh \Theta_{sc} - \Theta_{sc}\right)^{2/3}\over 5/3}
      \approx -2.81\,. 
\end{equation}
In terms of the primordial density field, the shell-crossing 
redshift $z_{sc}$ of a void with (linearly extrapolated) density 
deficit $\Delta_{lin,0}$ may therefore be directly predicted.  
For an Einstein-de Sitter Universe it is  
\begin{equation}
1+z_{sc}\,=\,{\displaystyle |\Delta_{lin,0}|\over 2.8059}\,.
\end{equation}
And at shell-crossing, the void has a precisely determined 
excess Hubble expansion rate:
\begin{equation}
 H_s = (4/3)\,H_u(t_{sc}),
\end{equation}
with $H_u=H_u(t_{sc})$ the global Hubble expansion factor at $t_{sc}$.  

For a spherical underdensity, the instant of shell crossing marks a dynamical 
phase transition. It is as significant as the full collapse stage reached by 
an equivalent overdensity. Also, as with the collapse of the overdensity 
the timescales on which this happens are intimately related to the 
initial density of the perturbation. The instant of shell crossing is 
determined by the global density parameter $\Omega_i$, 
the initial density deficit $\Delta_i$ of the shell, 
and the steepness of the density profile. 
In turn, this link between the initial void configuration and the 
void's {\it shell crossing} transition epoch paves the way towards 
predicting the nonlinear evolution of the cosmic void population on 
the basis of the primordial density field. 

\subsection{Beyond shell-crossing}
Virtually all early studies of void evolution concentrated on 
analytically tractable configurations of symmetric holes in a uniform 
background, either with or without compensating ridges. 
This allowed Hoffman \& Shaham~\cite{hoffshah82} to argue that voids 
should indeed be a seen as a natural outcome of a dissipationless 
clustering scenario, evolving from deep underdense regions in the 
primordial density field. This was followed by a variety of similar 
numerical studies (Peebles~1982;  
Hoffman, Salpeter \& Wasserman~1983; Hausman, Olson \& Roth~1983). 
The most extensive and systematic study of spherical void evolution, 
the work by Bertschinger~(1983, 1985) for voids 
in an Einstein--De~Sitter universe, concluded that in most viable 
circumstances voids would develop a dense surrounding shell. 
Following shell crossing at these void boundaries, the void would 
enter a phase of nonlinear evolution characterized by a self-similar 
outward expansion (also see Suto et al.~1984; 
Fillmore \& Goldreich~1984). On the basis of this, 
Blumenthal et al.~\cite{bdglp92} attempted to relate the derived 
void characteristics to the observed galaxy distribution.  
Dubinski et al.~\cite{ddglp93} (also see Van de Weygaert \& Van Kampen~1993) 
showed that when this was done, then the spherical tophat model 
provided a rather good description of void formation and evolution 
in their numerical simulations. 
The spherical model is equally succesfull in describing the evolution 
of spherical voids with more generic density profiles, and can be 
employed to demonstrate that they will often quickly evolve towards 
a tophat configuration (see Fig.~\ref{sphmodel}). 
Therefore, a description of void evolution which is based on the 
spherical evolution model, a strategy which we will follow in the 
main text, is amply justified.

\subsection{Beyond the spherical model}
We have concentrated on the evolution of spherical perturbations.  
However, generic peaks in Gaussian random fields are triaxial 
(BBKS~1986), so it is worth spending a little time discussing 
the evolution of ellipsoidal perturbations.  
It turns out that, as the underdense ellipsoid expands, the spherical 
model becomes an increasingly good approximation. 

A simple approximation for the gravitational potential in the 
immediate vicinity of a density minimum is a second order scheme, 
which approximates the density field by an ellipsoid of uniform density.  
The evolution of {\it low\/}-density regions may therefore be 
approximated via the equations of motion for a 
{\em homogeneous ellipsoid}. 
The description of a void's evolution is therefore analogous to the 
equivalent description of the collapse of overdensities 
(Icke~1972,~1973; White \& Silk~1979). 
Bond \& Myers~\cite{bm96} noted that it is possible to incorporate 
external (anisotropic) influences through the appropriate modification 
of the equation of motion (this same scheme was adopted by 
Eisenstein \& Loeb~1995). 

In the case of overdense regions, from which collapsed halos form, 
the ensuing nonlinear evolution tends to strongly amplify these 
initial departures from sphericity (Lin, Mestel \& Shu~1965). 
The collapse of overdensities typically proceeds in an anisotropic 
fashion, progressing through an increasingly flattened and elongated 
configuration before the ultimate collapse along all directions is 
complete. 
The key towards understanding this tendency is the anisotropic force 
field corresponding to nonspherical objects. In the case of an 
overdensity, the effective gravitational forces are directed inward, 
which, in combination with their anisotropy, translates into an 
increased rate of collapse along the shortest axis. 
In the cosmological context, this explains the existence of 
filamentary and sheetlike structures on Megaparsec scales 
(Icke~1973; White \& Silk~1979).  

On the basis of the same arguments, voids become increasingly spherical 
as they evolve (Icke~1984; Bertschinger~1985). 
That is, the anisotropic peculiar force field directed outward will 
induce the {\it strongest} acceleration along the {\it shortest axis}, 
causing the void to {\it expand fastest} along that direction. 
In contrast, a weaker acceleration along the longest axis leads 
to a smaller rate of excess expansion. Hence, the tendency of 
underdense regions to {\it nullify initial asphericities} and 
evolve into an {\it ever more spherical} geometry. 
Moreover, for a broad range of initial density profiles, voids will 
develop into objects with a distinctly tophat configurations. 
The reason for this is the same as for the spherical underdensities 
studied above.  This evolution towards a tophat profile was indeed 
observed by Van de Weygaert \& Van Kampen~\cite{vdwvk93} for voids 
in more generic circumstances.  The homogeneous interior density goes 
along with a uniform velocity divergence. Thus, generic primordial 
underdensities appear to evolve into ``super-Hubble expanding bubbles'' 
(Icke~1984; Van de Weygaert \& Van Kampen~1993). 

Of course the ellipsoidal model has serious limitations. 
It disregards important aspects like the presence of substructure. 
More serious is its neglect of any external influence, whether 
secondary infall, ``collision'' with surrounding matter (neighbouring 
expanding voids!), or the role of nonlocal tidal fields. 
Yet, it is interesting that in the case of voids the homogeneous 
ellipsoidal model becomes a better approximation over an ever 
increasing volume of space, as time proceeds. 
This has been confirmed by N-body simulatios of void evolution in 
realistic clustering scenarios which show how the matter distribution 
in the central region of (proto)voids flattens out as they expand and 
get drained (e.g. Van de Weygaert \& Van Kampen~1993, Fig. 31). 
On the basis of the spherical model (see e.g. Fig.~\ref{sphmodel}) 
one may readily appreciate this generic {\it flattening} of the density 
profile and its outwards {\it expansion}. Thus, we have the ellipsoidal 
model providing the argument for the sphericity of voids, and the spherical 
model demonstrating why the required conditions for the applicability 
of the ellipsoidal model are generically encountered in the case of 
voids. 

\section{Random field characteristics}\label{grfs}
When evaluating the statistics of a three dimensional random field 
of density perturbations filtered on a specific spatial scale $R$, 
the spectral moments, 
\begin{equation}
 \sigma^2_j(R)\,=\,\int {{\rm d}k\over k}\,{k^3\,P(k)\over 2\pi^2}\, 
                        k^{2j}\, |{\hat W}(kR)|^2 \,,
 \label{sigmaj}
\end{equation}
play a key role.  
Here $P(k)$ denotes the power spectrum of the unsmoothed density 
initial fluctuation field, extrapolated using linear theory to the 
present time, and ${\hat W}(kR)$ represents the shape of the filter. 
For example, if the density field is smoothed with a tophat or 
Gaussian filter, then ${\hat W}(x)$ is 
\begin{eqnarray}
 {\hat W}_{\rm TH}(x) & = & (3/x^3)\,(\sin x - x\cos x)\,,\\ 
 {\hat W}_{\rm G}(x)  & = & \exp(-x^2/2)\,,
\end{eqnarray}
respectively.  The total volume enclosed by these filters is 
$V_{\rm TH}=4\pi R^3/3$ and $V_{\rm G}=(2\pi)^{3/2}R^3$ respectively.  
The mass within the filter is $m=\bar\rho V(1+\delta)$, where 
$\delta$ represents the overdensity:  it is this quantity which 
fluctuates from one position to another.  If the density fluctuations 
are small everywhere, then the mass within a filter is approximately 
the same everywhere:  $m\approx\bar\rho V$.  

In models of hierarchical structure formation, the initial 
fluctuations around the mean density are indeed small.  Therefore, 
the correspondence between mass and filter scale $m\propto R^3$ 
suggests that if one wishes to model (proto)objects of mass $m$, 
one should study the initial density fluctuation field when it is 
smoothed on (comoving) spatial scale $R\propto m^{1/3}$, with 
the exact coefficient depending on filter choice.  

Thus, we may analyze any fluctuation field quantity in terms of its 
spatial scale $R$, or mass scale $m$.  To illustrate, consider a 
power-law power spectrum $P(k)$, 
\begin{equation}
 P(k)\,\propto k^n\,,
\end{equation}
for which 
\begin{equation}
 \sigma^2_j(m)\,\propto\,m^{-(n+3+2j)/3}.  
\end{equation}
Notice that if  $-3<n\le 1$, then $\sigma_0(m)$ is a decreasing 
function of $m$.  This remains true for any spectrum whose 
``generalized'' power spectrum slope, 
\begin{equation}
n(k)\,=\,{\displaystyle d \log P(k) \over \displaystyle d \log k}\,,
\end{equation}
is within the range $-3<n(k)\le 1$, even if it is not necessarily 
constant over the whole spectral range. 

The quantity $\sigma_0(m)$ quantifies the root mean square amplitude 
of density fluctuations on mass scale $m$.  
It will feature prominently in this work.
In hierarchical scenarios, $\sigma_0(m)$ is a monotonously decreasing 
function of scale, so it will serve as a ``dimensionless'' parameter 
which characterizes and the ``scale'' of density fluctuations.  
We will often use $S(m)$ to denote $\sigma^2(m)$.  

To evaluate the number density of peaks of a scale $m$ in the initial  
Gaussian density field, one must take into account the shape of the 
power spectrum, through the spectral parameters $R_*$ and $\gamma$, 
\begin{equation}
 R_*\,=\,\sqrt{3}\,{\displaystyle \sigma_1 \over \displaystyle \sigma_2}; 
 \qquad 
 \gamma\,=\, {\displaystyle \sigma_1^2 \over \displaystyle \sigma_0 \sigma_2} 
 \label{bbksrg}
\end{equation}
where $\sigma_0$, $\sigma_1$ and $\sigma_2$, which depend on the 
shape of the power spectrum, are defined by equation~(\ref{sigmaj}).  

\subsection{Simple Peaks Model}\label{peaks}
The number density of peaks (minima) of height (depth) $\delta_p$  
in an initial Gaussian density fluctuation field, smoothed with a 
filter scale $R$ and corresponding mass scale  $m={\bar\rho} V_f$, 
has been been worked out by BBKS. They showed that the density 
of peaks (minima) of height (depth)
\begin{equation}
 \nu\ =\ {\displaystyle \delta^2_{\rm p} \over \displaystyle \sigma^2_0(m)}\,, 
\end{equation}
in a Gaussian random field is 
\begin{equation}
 n_{\rm v}(\nu)\,{\rm d}\nu = \sqrt{\nu\over 2\pi}\,
                              {\exp(-\nu/2)\over (2\pi)^{3/2} R_*^3}\,
                 {G(\gamma,\gamma\nu^{1/2})\over 2}\,{{\rm d}\nu\over\nu},
 \label{npkbbksapp}
\end{equation}
with the spectral parameters $R_*$ and $\gamma$ given in 
Eqn.~\ref{bbksrg}. For a power-law $P(k)$, $P(k)\propto k^n$, some of 
these integrals diverge if one uses a tophat filter. Without loss of 
physical meaning, it is therefore preferrable to use a Gaussian filter. 
The function $G$ is then given by
\begin{displaymath}
 G(\gamma,y) = \int_0^\infty {\rm d}x\,f(x)\,
             {\exp[-(x-y)^2/2(1-\gamma^2)]\over\sqrt{2\pi (1-\gamma^2)}},
\end{displaymath}
in which the function $f(x)$ is defined as 
\begin{eqnarray*}
f(x) &=& {(x^3 - 3x)\over 2}
 \left\{{\rm erf}\left[\left(5\over 2\right)^{1/2}x\right] + {\rm
 erf}\left[\left(5\over 2\right)^{1/2}{x\over 2}\right]\right\} \nonumber \\
 && + \left(2\over 5\pi\right)^{1/2} 
     \Biggl[\left({31x^2\over 4}+{8\over 5}\right)\exp(-5x^2/8) \nonumber \\
 && \qquad\qquad\qquad + 
     \left({x^2\over 2} - {8\over 5}\right)\exp(-5x^2/2)\Biggr].
 \label{bbksfx}
\end{eqnarray*}

\section{First crossings and linear barriers}\label{fs2bar}
Let $f(x)$ denote the first crossing distribution of a single barrier 
of constant height $B$:
\begin{equation}
 f(x,B)\,{\rm d}x = \left({B^2\over 2\pi x}\right)^{1/2}\,
                    \exp\left(-{B^2\over 2x}\right)\,{{\rm d}x\over x}.
\end{equation}
The Laplace Transform of this distribution is 
\begin{equation}
 L(t,B) = \int_0^\infty {\rm d}x\,f(x,B)\,{\rm e}^{-tx} 
        = \exp\left(-\sqrt{2tB^2}\right).
 \label{Ltconstant}
\end{equation}
If we set $S$ equal to $\sigma^2_0(m)$ defined by equation~(\ref{sigmaj}) 
then the distribution $f(S,\delta_{\rm c})$ gives the excursion set 
approach's approximation for the fraction of mass which is bound up in 
collapsed objects of mass $m(S)$ (Bond et al. 1991).  
Therefore, if $n(m)$ denotes the number density of such collapsed 
halos, then 
\begin{equation}
 {m^2\,n(m)\over\bar\rho} \equiv 
 S\,f(S,\delta_{\rm c}) \,{{\rm d\ ln}\,S\over {\rm d\ ln}\,m},
\end{equation}
where $\bar\rho$ is the background density.  
In models of hierarchical clustering, $S$ decreases monotonically 
with increasing $m$.  If the initial spectrum of fluctuations was a 
power law, then: $\delta_{\rm c}^2/S\propto (m/m_*)^{(n+3)/3}$ 
with $-3< n \le 1$.  

By analogy, the fraction of mass in voids which each contain mass 
$m(S)$ is given by 
\begin{equation}
 {\cal F}(S,\delta_{\rm v},\delta_{\rm c}) = f(S,\delta_{\rm v}) - 
 \int_0^S\! {\rm d}s\,{\cal F}(s,\delta_{\rm c},\delta_{\rm v})\,
            f(S,\delta_{\rm v}|s,\delta_{\rm c}):
 \label{fsconvolution}
\end{equation}
the first term on the right hand side is the first crossing distribution 
of the barrier $\delta_{\rm v}$, and the second term subtracts from it 
the subset of trajectories which had crossed $\delta_{\rm c}$ before 
ever reaching $\delta_{\rm v}$.  
Since $f(S,\delta_{\rm v}|s,\delta_{\rm c}) 
       = f(S-s,\delta_{\rm c}-\delta_{\rm v})$, the Laplace Transform 
of ${\cal F}$ is 
\begin{eqnarray}
 {\cal L}(t,\delta_{\rm v},\delta_{\rm c}) &=& L(t,\delta_{\rm v}) - 
    \int_0^\infty\!\! {\rm d}s\,{\cal F}(s,\delta_{\rm c},\delta_{\rm v})\, 
                                {\rm e}^{-ts}\nonumber\\
 && \qquad \quad \times \int_{S-s}^\infty\!\! {\rm d}S\,
           f(S-s,\delta_{\rm c}-\delta_{\rm v})\,{\rm e}^{-t(S-s)}\nonumber\\
 &=& L(t,\delta_{\rm v})- {\cal L}(t,\delta_{\rm c},\delta_{\rm v})\,
                         L(t,\delta_{\rm c}-\delta_{\rm v}).
 \label{ltvc}
\end{eqnarray}
Whereas the actual distributions are related by convolutions 
(equation~\ref{fsconvolution}), the Laplace transforms simply multiply: 
in this respect, the Laplace transforms behave similarly to one's 
intuition about the independence of walks to and between the barriers.  
By symmetry, 
\begin{equation}
 {\cal L}(t,\delta_{\rm c},\delta_{\rm v}) = 
  L(t,\delta_{\rm c})- {\cal L}(t,\delta_{\rm v},\delta_{\rm c})\,
                         L(t,\delta_{\rm v}-\delta_{\rm c}).
 \label{ltcv}
\end{equation}
Inserting this into equation~(\ref{ltvc}) yields 
\begin{eqnarray}
 {\cal L}(t,\delta_{\rm v},\delta_{\rm c}) &=& 
 {L(t,\delta_{\rm v}) - L(t,\delta_{\rm c})\,L(t,\delta_{\rm c}-\delta_{\rm v})
\over 1-L(t,\delta_{\rm v}-\delta_{\rm c})\,L(t,\delta_{\rm c}-\delta_{\rm v})}
\nonumber \\
 &=& {\sinh[\delta_{\rm c}\sqrt{2t}]\over 
      \sinh[(\delta_{\rm c}-\delta_{\rm v})\sqrt{2t}]}.
 \label{ltconstant}
\end{eqnarray}
Inverting this Laplace Transform yields equation~(\ref{vfvoid}) 
in the main text.  

Notice that 
\begin{equation}
 L(0,\delta_{\rm c}) = 1 \qquad {\rm and} \qquad
 {\cal L}(0,\delta_{\rm v},\delta_{\rm c}) = 
 {\delta_{\rm c}\over \delta_{\rm c}-\delta_{\rm v}}.  
 \label{lt0}
\end{equation}
The first equality shows that all random walks cross $\delta_{\rm c}$, 
and is interpretted as indicating that all mass is associated with 
gravitationally bound halos.  In constrast, the second equality shows 
that only a fraction of all random walks cross $\delta_{\rm v}$ without 
first having crossed $\delta_{\rm c}$; evidently only a fraction 
$f_{\rm void} = \delta_{\rm c}/(\delta_{\rm c}-\delta_{\rm v})$ of the 
mass is associated with voids.  

Although we do not use this fact in the main text, the calculation 
above can be generalized to include barriers of the form 
$B_{\rm c} = \delta_{\rm c} - \beta S$.  
These are barriers which are not constant; rather, their height decreases 
linearly with $S$ if $\beta > 0$, and increases if $\beta$ is negative.  
The first crossing distribution of linear barriers is Inverse Gaussian; 
the associated Laplace transform is 
$\exp[\beta\delta_{\rm c} - \sqrt{\delta_{\rm c}^2(2t+\beta^2)}]$ 
(e.g. Sheth 1998); it reduces to equation~(\ref{Ltconstant}) when 
$\beta=0$.  

If both barriers change linearly with $S$, but they have the same slope, 
$B_{\rm c} = \delta_{\rm c} - \beta S$ and 
$B_{\rm v} = \delta_{\rm v} - \beta S$, then 
exactly the same reasoning which led to equation~(\ref{ltconstant}) 
yields
\begin{equation}
 {\cal L}(t,\delta_{\rm v},\delta_{\rm c},\beta) = 
  \exp(\delta_{\rm v}\beta)\,{\sinh[\delta_{\rm c}\sqrt{\beta^2 + 2t}]\over 
                  \sinh[(\delta_{\rm c}-\delta_{\rm v})\sqrt{\beta^2 + 2t}]}
 \label{ltlinear}
\end{equation}
for the Laplace transform of the distribution which crosses 
$B_{\rm v}$ without first crossing $B_{\rm c}$.  
(The first crossing of $\delta_{\rm c}$ without crossing $\delta_{\rm v}$ 
is given by interchanging `c' and `v' in the expression above.)  
Note that ${\cal L}(0) = \exp(\delta_{\rm v}\beta)\,
  \sinh(\delta_{\rm c}\beta)/\sinh[(\delta_{\rm c}-\delta_{\rm v})\beta]$; 
as in the constant barrier model, only a fraction of walks cross 
one barrier without first crossing the other.  
Inverting this Laplace transform yields  
\begin{eqnarray}
 S\,{\cal F}(S,\delta_{\rm v},\delta_{\rm c}) &=& 
    {\rm e}^{\beta\delta_{\rm v}}\sum_{j=1}^\infty {(j\pi{\cal D})^2\over\nu}\,
            {\sin(j\pi {\cal D})\over j\pi}  \nonumber\\
 && \qquad \ \times 
           \exp\left(-{(j\pi {\cal D})^2 + \beta^2\delta_{\rm v}^2\over 2\nu}\right),
 \label{vfvlinear}
\end{eqnarray}
where $\nu\equiv \delta_{\rm v}^2/S$ and 
${\cal D}=|\delta{\rm v}|/(\delta_{\rm c}-\delta_{\rm v})$.  


\end{document}